\documentclass[fleqn,usenatbib]{mnras}

\usepackage{newtxtext,newtxmath}

\usepackage[T1]{fontenc}
\usepackage{ae,aecompl}

\usepackage{graphicx}    
\usepackage{amsmath}    
\usepackage{amssymb}    
\usepackage{amssymb}    
\usepackage{amsbsy}    

\newcommand{\Dnu}{\Delta_\ell}

\newcommand{\Teff}{T_{\rm{eff}}}

\defcitealias{Joergensen2017}{J17}
\defcitealias{Joergensen2018}{J18}
\defcitealias{JoergensenWeiss2019}{JW19}

\title[MCMC with coupled 1D and 3D stellar models]{Bayesian inference of stellar parameters based on 1D stellar models coupled with 3D envelopes}

\author[J{\o}rgensen and Angelou]{
Andreas Christ S{\o}lvsten J{\o}rgensen$^{1}$\thanks{E-mail: acsj@mpa-garching.mpg.de} and
George C. Angelou$^{1,2}$ 
\\
$^{1}$Max-Planck-Institut f\"ur Astrophysik, Karl-Schwarzschild-Str. 1, D-85748 Garching, Germany \\
$^{2}$Max-Planck-Institut f\"{u}r Sonnensystemforschung, Justus-von-Liebig-Weg 3, 37077 G\"{o}ttingen, Germany \\
}

\date{Accepted XXX. Received YYY; in original form 29.05.2019}

\pubyear{2019}

\begin{document}
\label{firstpage}
\pagerange{\pageref{firstpage}--\pageref{lastpage}}
\maketitle

\begin{abstract}
{{Stellar models utilising one-dimensional (1D), heuristic theories of convection fail to adequately describe the {\color{black}energy} transport in superadiabatic layers. The improper modelling leads to well-known {\color{black}discrepancies} between observed and predicted oscillation frequencies for stars with convective envelopes. Recently, three-dimensional (3D) hydrodynamic simulations of stellar envelopes have been shown to facilitate a realistic depiction of superadiabatic convection in 1D stellar models. The resulting structural changes of the boundary layers have been demonstrated to impact not only the predicted oscillation spectra but evolution tracks as well. In this paper, we quantify the consequences that the change in boundary conditions has for stellar parameter estimates of main-sequence stars. For this purpose, we investigate two benchmark stars, Alpha Centauri A and B, using Bayesian inference. We show that the improved treatment of turbulent convection makes the obtained 1D stellar structures nearly insensitive to the mixing length parameter. By using 3D simulations in 1D stellar models, we hence overcome the degeneracy between the mixing length parameter and other stellar parameters. 
{\color{black}By lifting this degeneracy, the inclusion of 3D simulations has the potential to yield} more robust parameter estimates. {\color{black}In this way, a more realistic depiction of superadiabatic convection has} important implications for any field that relies on stellar models, including the study of the chemical evolution of the Milky Way Galaxy and exoplanet research.}}

\end{abstract}




\section{Introduction}

Parameterizations of superadiabatic convection, such as mixing length theory \citep[MLT,][]{Bohm-Vitense1958} or full-spectrum turbulence theory \citep{Canuto1991,Canuto1992}, do not fully capture the complexity of convection. Since convection plays a key role in stellar structures and their evolution, the use of these {\color{black}simplifying approximations} in stellar evolution codes leads to notable shortcomings of the obtained stellar models. 

For instance, for stars with convective envelopes, the incorrect depiction of the outer boundary layers is known to lead to systematic errors in the predicted model frequencies. This shortcoming is known as the structural surface effect. The surface effect also has a modal contribution \citep[cf. ][]{Houdek2017}. {This contribution stems from the assumption of adiabaticity in the frequency calculations. It is thus not related to the evaluated stellar structure but to the frequency calculations. The combined surface effect} is commonly addressed in the post-processing by using empirical correction relations \citep{Kjeldsen2008,Ball2014,Sonoi2015}. Alternatively, one may substitute the outermost layers with more realistic envelopes in the post-processing, based on multi-dimensional hydrostatic simulations of convection. This approach is known as patching (\citealt{Rosenthal1999,Piau2014,Sonoi2015, Ball2016, Magic2016, Joergensen2017} [\citetalias{Joergensen2017}], \citealt{Trampedach2017,Joergensen2019}).

For stars with convective envelopes, the simplified treatment of superadiabatic convection also provides incorrect boundary conditions, which affects the solution of the stellar structure equations. As shown by \cite{Mosumgaard2018} {\color{black}and \cite{JoergensenWeiss2019} [\citetalias{JoergensenWeiss2019}]}, the use of MLT hence alters the evolution tracks of stellar models. This is not accounted for when patching, i.e. when only correcting for the use of MLT in the post-processing. 

Recently, \citet{Joergensen2018} [\citetalias{Joergensen2018}] have, therefore, proposed a novel method that ensures a realistic depiction of convective surface layers, throughout the stellar evolution. They do this by employing 3D simulations of stellar envelopes: at each time-step of the evolution, the mean stratification of a 3D simulation is appended and is used to set the outer boundary conditions.
The method corrects for the structural surface effect and supplies realistic boundary conditions, whereby the evolution tracks have been shown to be altered (cf. \citetalias{JoergensenWeiss2019}). In the following, we will refer to this method as the \textit{coupling of 1D and 3D simulations} {\color{black}(cf. Section~\ref{sec:coupling})}.

Due to the effect of the improved boundary layers on the predicted evolution tracks, the coupling of 1D and 3D models may alter stellar parameter estimates inferred from spectroscopic and asteroseismic constraints. Thus, the treatment of superadiabatic convection in stellar evolution codes has ramifications for any field that relies on the accuracy of stellar models. This includes exoplanet research, through the characterization of the host stars, and galactic archaeology, through the ages and chemical composition that is attributed to stellar populations. Realistic stellar models are furthermore crucial for studying physical processes in stars.

In this paper, we explore the implications of the coupling of 1D and 3D models for the evaluated stellar parameters of main-sequence {\color{black}(MS)} stars. For this purpose, we employ a Markov Chain Monte Carlo (MCMC) approach. Monte Carlo methods, such as MCMC, allow for a robust exploration of the parameter space and yield reliable probability distributions for target parameters. 
A variety of such sampling schemes have thus entered all fields of astronomy, reaching from the interpretation of \textit{Kepler} light-curves to cosmology \citep[e.g.][]{Handberg2011,Lund2017,Jasche2013,Porqueres2019}. 
However, due to the high computational cost, MCMC is seldom used in order to determine stellar properties with some notable exceptions, including the endeavour to obtain a detailed characterization of the present-day Sun\footnote{For some purposes, however, it is worth noting that alternative approaches with low computational cost are able to recover the correct posterior probability distributions. Thus, the posterior probability distribution of the solar neutrino flux can be derived semi-analytically, due to the linear response of the flux to changes in the stellar parameters \citep{Joergensenjcd2017}.} \citep{Bahcall2006,Vinyoles2017} and its closest neighbours, Alpha Centauri A and B \citep{Bazot2012}. This binary is likewise the target that we have decided to investigate in detail in this paper. { Crucially, the orbital analysis \citep{Pourbaix2016} provides the necessary independent mass prior for an otherwise wide and correlated parameter space}. Due to its history as a benchmark for stellar physics \citep[e.g.][]{2018ApJ...864...99J, 2019MNRAS.tmp.2125S} and the high-quality data that is available for this binary, it is an ideal target for a detailed differential comparison between models with different boundary conditions. 

\section{Stellar Models}

In this paper, we employ the Garching stellar evolution code \citep[\textsc{garstec,}][]{Weiss2008}. We compute stellar structures and their evolution, following the approach proposed by \citetalias{Joergensen2018}: at every time-step of the evolution, the average stratification of an interpolated 3D stellar envelope is appended to the stellar model and used to set the boundary conditions of the stellar interior. We will refer to these models as \textit{coupled stellar models}. 

{The fact that our coupled models include the full mean stratification of 3D simulations throughout their evolution makes the employed coupling scheme novel. No other scheme includes such detailed information from 3D simulations. {\color{black}In stead, alternative approaches} either rely on parameterizations of the 3D simulations or {\color{black}on patching such simulations onto standard stellar models that employ MLT or similar approximations to determine the outer} boundary conditions. (e.g. \citealt{Rosenthal1999,Piau2014,Trampedach2014b,Sonoi2015,Ball2016},\citetalias{Joergensen2017},\citealt{Trampedach2017,Mosumgaard2018,2018ApJ...869..135S}).}

The method by \citetalias{Joergensen2018} relies on the interpolation scheme by \citetalias{Joergensen2017}, in order to compute the appended mean 3D envelope --- to which we will refer as the $\langle \mathrm{3D} \rangle$-envelope. The method has been shown to recover the mean stratification of 3D hydrodynamic simulations accurately. Thus, the appended $\langle \mathrm{3D} \rangle$-envelope ensures that the correct boundary conditions are supplied to the stellar interior and accurately mimics the properties of multi-dimensional hydrodynamic simulations.

The interpolation scheme by \citetalias{Joergensen2017} allows for an interpolation in effective temperature ($T_\mathrm{eff}$) and gravitational acceleration ($g$). Recently, \cite{Joergensen2019} have expanded upon this scheme allowing for interpolation in metallicity ($\mathrm{[Fe/H]}$). Building upon this work, we append 3D envelopes at every time-step of the evolution, interpolating in a three-dimensional parameter space. However, in our models, we ignore the diffusion of elements. We consequently assume the metallicity to be fixed, throughout the evolution. Moreover, for each star, we fix $\mathrm{[Fe/H]}$ in our Bayesian analysis: we vary the hydrogen abundance ($X$), the helium abundance ($Y$), and the abundance of heavy elements ($Z$) but keep {\color{black}the} ratio of $Z$ to $X$ fixed to a predetermined value for every model in the Markov chain. 


\subsection{Coupling 1D and 3D models} \label{sec:coupling}

Stellar evolution codes solve a set of five differential equations, the so-called stellar structure equations, in order to evaluate stellar structures and their evolution. At every time-step, an equilibrium structure is computed based on four of these equations, assuming a fixed chemical composition. The chemical profile is then evolved in time, keeping the stratification of the thermodynamic quantities fixed \citep[e.g.][]{Weiss2008,Kippenhahn}.

To obtain an equilibrium structure, i.e. to solve the stellar structure equations, suitable inner and outer boundary conditions must be supplied. The outer boundary conditions are commonly supplied by integration over theoretical model atmospheres, such as an Eddington grey atmosphere \citep[cf.][]{Kippenhahn}. These model atmospheres are {\color{black}often} analytic expressions, giving an approximation for the relation between the temperature ($T$) and the optical depth ($\tau$). They are hence referred to as $T(\tau)$ relations.
In standard stellar models that employ $T(\tau)$ relations, the outer boundary conditions are typically set at the photosphere, i.e. within the superadiabatic surface layers.

In our coupled stellar models, on the other hand, we supply the outer boundary conditions based on interpolated $\langle \mathrm{3D} \rangle$-envelopes rather than using $T(\tau)$ relations. Furthermore, we partly supply the outer boundary conditions at the base of the interpolated $\langle \mathrm{3D} \rangle$-envelopes, deep within the nearly-adiabatic region. Specifically, at the base of the envelope, we require that the model is continuous. In other words, we require that there are no discontinuities at the interface between the interior model and the imposed $\langle \mathrm{3D} \rangle$-envelope.

Beyond the outer boundary of the interior model, we append {\color{black}an interpolated} $\langle \mathrm{3D} \rangle$-envelope at every time-step. Following the terminology introduced by \citetalias{Joergensen2018}, we refer to the lowermost point of the appended interpolated $\langle \mathrm{3D} \rangle$-envelope as the matching point.

In addition to the already mentioned requirement of achieving a continuous stratification at the matching point, we require that the energy flux at the matching point is consistent with the effective temperature of the 3D simulation {\color{black}in question}. This outer boundary condition amounts to the requirement that the Stefan-Boltzmann law is fulfilled at the photosphere of the resulting hybrid model.

Most stellar evolution codes, including \textsc{garstec}, solve the stellar structure equations and impose the outer boundary conditions using the so-called Henyey-scheme \citep[cf.][]{Henyey1964,Kippenhahn}. Within this scheme, the interior structure, as well as the outer boundary conditions, are iteratively adjusted at every time-step, until an equilibrium structure is determined. The chemical profile is then evolved in time. For our coupled models, this implies that we adjust the interior structure and compute a new interpolated $\langle \mathrm{3D} \rangle$-envelope in each of these iterations at every time-step, to achieve consistent coupled equilibrium structures.

\begin{figure}
\centering
\includegraphics[width=0.85\linewidth]{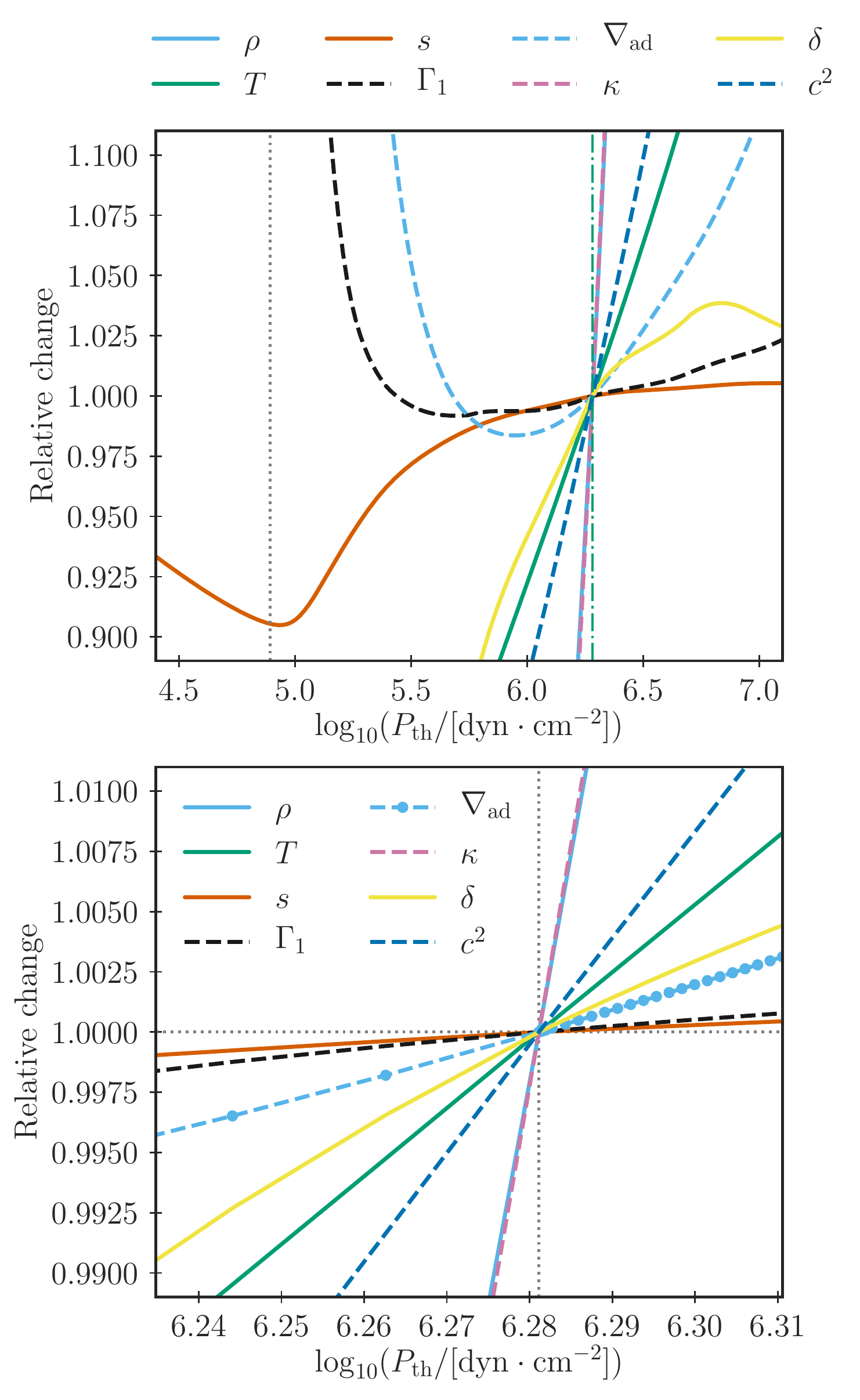}
\caption{\color{black}\textbf{Upper panel:} Structure of the uppermost layers of a coupled solar model, showing the change in several physical quantities relative to the corresponding value taken at the matching point as a function of the thermal pressure ($P_\mathrm{th}$). We include the density ($\rho$), the temperature ($T$), the first adiabatic index ($\Gamma_1$), the entropy ($s$), the adiabatic temperature gradient ($\nabla_\mathrm{ad}$), the Rosseland mean opacity ($\kappa$), the derivative of density with respect to temperature at constant thermal pressure ($\delta=-\partial \ln \rho/\partial \ln T$), and the squared sound speed ($c^2$). The dash-dotted green line shows the location of the matching point. The dotted grey line indicates the location of the photosphere. \textbf{Lower panel:} Zoom-in on the region around the matching point to illustrate the continuous transition of our models. The dots along the line for $\nabla_\mathrm{ad}$ indicate the location of the different mesh points in the coupled model. The intercept between the dotted grey lines indicates the location and value at the matching point.
}
\label{fig:sun_physqua}
\end{figure}

{\color{black}More specifically, as discussed in detail in \citetalias{Joergensen2018} and \citetalias{JoergensenWeiss2019}, we obtain the temperature stratification as a function of the thermal pressure for the outermost layers of coupled models based on existing $\langle \mathrm{3D} \rangle$-envelopes by interpolation in $\log T_\mathrm{eff}$ and $\log g$. We then require the temperature and thermal pressure at the base of the $\langle \mathrm{3D} \rangle$-envelope to be matched by the interior structure.
The remaining physical quantities, such as the density, are inferred by the stellar evolution code from the equation of state (EOS) and the opacity tables, throughout the appended envelope. All physical quantities are, therefore, by construction, continuous at the matching point. Furthermore, as shown by \citetalias{Joergensen2018} and \citetalias{JoergensenWeiss2019}, the hereby obtained structure reliably mimics the mean stratifications of the underlying 3D simulations --- this statement also applies to quantities, such as the density, that are inferred by the stellar evolution code.}
{\color{black}Figure~\ref{fig:sun_physqua} illustrates this by showing the stratification of several physical quantities, including several physically relevant derivatives, for a coupled structure model of the present-day Sun.}

{\color{black}Since the consistent structure of coupled models heavily relies on the imposed boundary conditions, it is worth elaborating on our implementation.}
{\color{black}The scheme presented by \citetalias{Joergensen2018} involves interpolation in two different planes: first, we interpolate in the $(\log T_\mathrm{m},\log g)$-plane, where $T_\mathrm{m}$ denotes the temperature at the matching point, to obtain a proposal for $T_\mathrm{eff}$. Subsequently, we interpolate in the $(\log T_\mathrm{eff},\log g)$-plane to obtain $T(P_\mathrm{th})$, where $P_\mathrm{th}$ denotes the thermal pressure. To ensure that the resulting temperature stratification is continuous, it must be guaranteed that the required interpolation in the $(\log T_\mathrm{eff},\log g)$-plane recovers the correct temperature at the matching point. In order to accomplish this, the interpolation in the $(\log T_\mathrm{m},\log g)$-plane is merely used as a stepping stone, based on which $T_\mathrm{eff}$ can be pinpointed. Thus, based on the interpolation in the $(\log T_\mathrm{m},\log g)$-plane, we compute $T_\mathrm{m}^\mathrm{3D}$ for different values of $T_\mathrm{eff}$ within $5\,\%$ of the suggested value.
We then adopt the value of $T_\mathrm{eff}$, for which the interpolation in the $(\log T_\mathrm{eff},\log g)$-plane minimizes the discontinuity in the temperature stratification at the matching point. For many models, including the solar model in Fig.~\ref{fig:sun_physqua}, a continuous temperature stratification can be ensured within the machine precision of the calculation. Thus, the computation of coupled models effectively only involves interpolation in one plane, the $(\log T_\mathrm{eff},\log g)$-plane. The continuous stratification in thermal pressure is ensured iteratively using the Henyey scheme. The continuous temperature stratification is ensured in each iteration of the Henyey scheme by following the strategy described above.}

The method for coupling 1D and 3D models by \citetalias{Joergensen2018} hence ensures a high level of consistency between the 1D stellar interior and the appended $\langle \mathrm{3D} \rangle$-envelope: not only does the stratification of the 1D interiors of the coupled models match all quantities at the base of the corresponding $\langle \mathrm{3D} \rangle$-envelopes, but this match is ensured at every single time-step.

Instead of taking information from 3D simulations directly into account throughout the computed stellar evolution, other authors only use 3D simulations to adjust the stellar structure at a given final age. The method presented by \cite{Trampedach2014a,Trampedach2014b} is a prominent exception to this rule. \cite{Trampedach2014a} have extracted a $T(\tau)$ relation for the atmospheric stratification above the photosphere based on 3D simulations --- as well as necessary modifications of the radiative temperature gradient down to an optical depth of 10, in order to properly depict the photospheric transition. Furthermore, \cite{Trampedach2014b} have determined the value of the mixing length parameter that must be used in 1D models to recover the temperature and pressure at the base of 3D simulations\footnote{{This calibration includes an additional free parameter associated with the turbulent pressure.}}. They find that this value varies throughout the Hertzsprung-Russell (HR) diagram. By including their calibrated variable mixing length parameter and the associated $T(\tau)$ relation into a stellar evolution code, the resulting 1D models are, therefore, able to mimic the underlying 3D simulations at and above the photosphere as well as at a single point in the nearly-adiabatic region. As shown by \citetalias{Joergensen2017} and \cite{Mosumgaard2018}, the resulting models do, however, not recover the correct stratification of the entire superadiabatic region: a single value of the mixing length parameter is insufficient to encode the complexity of these layers. 


\subsection{Standard stellar models}
For comparison, we also compute stellar models using MLT and plane-parallel Eddington grey atmospheres. In the following, we refer to such models as \textit{standard stellar models}. In contrast to the coupled stellar models, the outer boundary conditions of the standard stellar models are set at the photosphere.

Throughout this paper{, for standard models as well as for coupled models,} we use the FreeEOS by A.~W. Irwin \citep{Cassisi2003} and the OPAL opacities \citep{Iglesias1996}. We extend the latter by the opacities by \cite{Ferguson2005} at low temperatures. Furthermore, the composition is based on the solar composition determined by \cite{Asplund2009}, to which we will refer as AGSS09.


\section{The Stagger grid} \label{sec:stagger}

In this paper, we draw upon the 3D radiative hydrodynamical simulations of stellar convective envelopes by \cite{Magic2013}, when computing coupled stellar models. We will refer to this collection of simulations as the Stagger grid. 
{Simulations, such as the Stagger-grid {\color{black}calculations}, are often referred to as 3D atmospheres in the literature. They extend from the optically thin layers above the photosphere to the nearly-adiabatic interior. For comparison, Eddington grey atmospheres and other $T(\tau)$-relations truncate at the photosphere. In order to emphasize the larger depth of the 3D simulations, we hence refer to them as 3D \textit{envelopes}, following the nomenclature by \cite{Joergensen2018}. {\color{black} However, by no means do these simulations} cover the entire convective envelope. They only cover a small fraction of this region. Thus,} each simulation covers a representative volume of the surface, stretching from the nearly adiabatic interior to the stellar atmosphere, enclosing the photosphere. The extent of each envelope only corresponds to a small fraction of the total stellar radius \citep{Magic2013} and the gravitational acceleration is hence assumed to be constant throughout the envelope.

The Stagger grid spans simulations, for which the effective temperatures ($T_\mathrm{eff}$) lies between $4000\,$K and $7000\,$K in steps of $500\,$K, and for which the logarithm of the gravitational acceleration ($g$) lies between $1.5\,$dex and $5.0\,$dex in steps of $0.5\,$dex. The grid contains envelopes with a discrete set of metallicities: $\mathrm{[Fe/H]}$: $-4.0$, $-3.0$, $-2.0$, $-1.0$, $-0.5$, $0.0$ and $0.5$. For $\mathrm{[Fe/H]>-1.0}$, the relation between the metallicity and the mass fraction is
\begin{equation}
\mathrm{[Fe/H]}=\log_{10} \left(\frac{Z_\mathrm{S}}{X_\mathrm{S}} \right)- \log_{10}\left( \frac{\mathrm{Z}_{\mathrm{S},\odot}}{\mathrm{X}_{\mathrm{S},\odot}} \right).
\end{equation}
Here, $Z_\mathrm{S}$ and $X_\mathrm{S}$ denote the mass fraction of metals and hydrogen at the surface, respectively. $\mathrm{Z}_\mathrm{S,\odot}/\mathrm{X}_\mathrm{S,\odot}=0.1828$. For lower metallicities, $\alpha$-enhancement is accounted for, using $\mathrm{[\alpha/Fe]}=0.4$ \citep{Magic2013b}. We construct our coupled models in such a way that the metallicity of the interior model and the appended envelope are consistent.

The Stagger grid assumes the solar composition determined by AGSS09. For the sake of consistency, we hence use the same composition in our stellar structure models.
We note that the abundance of helium relative to hydrogen, i.e.
\begin{equation}
A(\mathrm{He})=\log_{10}\left( \frac{n_\mathrm{He}}{n_\mathrm{H}} \right)+12,
\end{equation}
is kept fixed, when varying the metallicity of the 3D simulations (cf. R.~collet, private communications).
Concretely, this implies that the helium mass fraction increases with decreasing metallicity, while the opposite behaviour would be more reasonable from {a cosmic chemical evolution perspective. The Stagger grid is hence by construction somewhat at odds with the expected chemical evolution. This is not unique to the Stagger grid; it also holds true for other sets of 1D and 3D envelope simulations}. In this paper, we ignore this inconsistency, and we allow the helium mass fraction to vary, as mentioned above.

{
\subsection{Interpolation scheme}

The interpolation scheme by \citetalias{Joergensen2017} {\color{black}for constructing $\langle \mathrm{3D} \rangle$-envelopes} builds upon the apparent homology of the mean stratification of 3D simulations: the mean stratifications of all 3D simulations look rather similar when scaled by the corresponding value at the minimum in $\partial\log\rho / \partial\log P_\mathrm{th}$ near the surface. Here, $\rho$ and $P_\mathrm{th}$ denote the density and the thermal pressure, respectively. For solar-like stars, this minimum corresponds to a plateau, while a density inversion may take place at this minimum in later evolutionary stages.

As shown by \citetalias{Joergensen2017}, it is possible to robustly recover the scaled structures of $\langle \mathrm{3D} \rangle$-envelopes as well as the associated scaling factors by interpolation. In our coupled models, this interpolation is performed directly by our stellar evolution code (cf. \citetalias{Joergensen2018}).

We note that we employ the thermal pressure as our coordinate, including both the gas pressure and the radiation pressure. Meanwhile, we ignore the turbulent pressure, since a consistent treatment of turbulent pressure in coupled stellar models has only recently been achieved by \citetalias{JoergensenWeiss2019} towards the end of the analysis that enters this paper. For our purposes, however, it is an acceptable approximation to neglect turbulent pressure: it is shown in 
\citetalias{JoergensenWeiss2019} that the implementation of turbulent pressure in coupled stellar models does not significantly alter the evolution tracks. Neither does turbulent pressure affect the stellar oscillation frequencies significantly, since we work within the gas $\Gamma_1$ approximation \citep{Rosenthal1999,Houdek2017}. In accordance with other authors \citep[e.g.][]{Sonoi2015,Ball2016,Magic2016}, we hence assume that the Lagrangian perturbation of the thermal pressure equals that of the turbulent pressure.}

Finally, we note that the base of the appended $\langle \mathrm{3D} \rangle$-envelope is placed at a pressure that is 16 times larger than the pressure at the minimum in $\partial\log\rho / \partial\log P_\mathrm{th}$ near the surface. This pressure is dictated by the shallowest simulation in the Stagger grid.


\section{Asteroseismic properties} \label{sec:astroprop}

We use the Aarhus adiabatic pulsation package, \textsc{adipls} \citep{Christensen-Dalsgaard2008a}, to compute stellar oscillation frequencies for our stellar structure models. These frequencies are used to compare the models to observations as specified in Section~\ref{sec:MCMC}

Because the frequencies are subject to the surface effect, a direct comparison without any corrections between the individual model frequencies and observations is uninsightful. 
For standard stellar structure models, this problem can be overcome by using surface correction relations.
However, when appending $\langle \mathrm{3D} \rangle$-envelopes, we partly mend the surface effect --- this is to say, modal effects, which includes non-adiabatic contributions, are still not corrected for \citep[e.g.][]{Houdek2017}. 
As a consequence, no suitable empirical surface correction relations exist. We can hence not compare standard stellar models with models that include 3D envelopes on an equal footing, by matching individual frequencies to observations. 
In stead, we use the frequency ratios suggested by \cite{Roxburgh2003}:
\begin{equation}
  r_{01}(n) = \frac{\nu_{n-1,0}-4\nu_{n-1,1} + 6 \nu_{n,0} - 4\nu_{n,1} + \nu_{n+1,0}}{8(\nu_{n,1}-\nu_{n-1,1})},
\end{equation}
\begin{equation}
  r_{10}(n) = \frac{-\nu_{n-1,1} + 4\nu_{n,0} - 6\nu_{n,1} + 4\nu_{n+1,0} - \nu_{n+1,1}}{8(\nu_{n+1,0}-\nu_{n,0})},
\end{equation}
\begin{equation}
  r_{02}(n) = \frac{\nu_{n,0}-\nu_{n-1,2}}{\nu_{n,1}-\nu_{n-1,1}}.
\end{equation}
Here, $\nu_{n,\ell}$ denotes the oscillation frequency of radial order $n$ and degree $\ell$. 
Analysis of the phase shifts \citep{Roxburgh2003} as well as kernels \citep{Oti2005} have demonstrated that the ratios $r_{01}$, $r_{10}$, and $r_{02}$ are relatively insensitive to the outermost layers. 

By coupling 1D and 3D models, we improve the physical depiction of the near-surface layers significantly.
Although it may thus seem as if the use of frequency ratios leads us to ignore exactly those structural features that we strive to improve, this is not the case: the outermost layers affect the entire interior structure and shift the evolution tracks. {\color{black}Although the ratios primarily probe the core region, they hence also reflect the improved boundary conditions.}
Furthermore, since the ratios mitigate the impact of the surface effect, they facilitate a comparison between model frequencies and data without the need for surface correction relations. 

In addition, we include large and small frequency separations in our comparison with data. The large separation is defined as follows:
\begin{equation}
\Delta_\ell = \nu_{n,\ell}-\nu_{n-1,\ell}. \label{eq:large}
\end{equation}
The small separations are defined by
\begin{equation}
d_{\ell \ell+2} = \nu_{n,\ell}-\nu_{n-1,\ell+2}.
\end{equation}
Although these two separations are impacted by the surface term, $\Dnu$, in particular, has been shown to be a far more sensitive diagnostic of stellar mass than the ratios \citep{2017ApJ...839..116A}. 
{\color{black}This being said, in the case of standard stellar models, we note that any model that recovers frequency separations without fully addressing the surface effect cannot correctly recover the individual oscillation frequencies. This is due to the fact that the surface effect leads to frequency residuals that increase with frequency, which, for instance, artificially increases the large separation that is inferred from the model frequencies. However, since we are merely interested in performing a differential study, we deem that the mentioned benefits outweigh this drawback. Thus,} the inclusion of the separations is a trade-off. They help constrain mass and age, and although they introduce some uncertainty, they display less sensitivity to the surface term than matching individual frequencies. {\color{black}Other authors, therefore, likewise include frequency separations in the likelihood when evaluating stellar parameters \citep[e.g.][]{Bazot2012}.}

{\color{black}Furthermore,} we note that all frequency combinations are correlated, both internally and mutually --- e.g. $r_{02}=d_{02}/\Delta_1$. These correlations must hence be taken properly into account in the statistical comparison between models and observations. We discuss this issue in Section~\ref{sec:MCMC}.


\section{Hephaestus, an optimization and search algorithm for stellar modelling} \label{sec:MCMC}

In this paper, we use the algorithm \textsc{hephaestus} (Angelou in prep., to which we refer for further details). In short  \textsc{hephaestus} is a stellar model optimization and search pipeline comprising genetic, downhill simplex and MCMC algorithms. Designed with the purpose of calculating full models on the fly, the code's API allows the user to straightforwardly graft any stable evolution or oscillation code. 
The genetic and downhill simplex algorithms aggressively identify local minima in a multidimensional stellar evolutionary parameter space (e.g., \citealt{2013ApJS..208....4P,2009ApJ...699..373M}) and have been modified to efficiently make use of parallel computational resources. 
Once these algorithms identify local minima, the model parameters can be passed to the MCMC algorithm to explore the posterior distribution in detail. 
The MCMC library also complements rapid search algorithms such as the Stellar Parameters in an Instant Pipeline (SPI, \citealt{2016ApJ...830...31B, 2017ApJ...839..116A}) which use regression to infer stellar parameters. 
SPI  distributions can serve as starting points for the walkers so that they may identify full stellar models for deeper analysis (e.g., inversions). 

The MCMC library is utilized to determine stellar parameters within the framework of Bayesian statistics. This means that
we ascribe probability as a measure of our belief in a proposition, based on data ($\boldsymbol{D}$) and prior information ($\boldsymbol{I}$).  
We will elaborate upon the concepts involved below, adopting the notation used by \cite{Gregory} and \cite{Handberg2011}.

Consider a model with $N$ parameters. These parameters are proposed to take the values $\boldsymbol{\Theta}$. 
We assume that these parameters are continuous, which implies that we will be dealing with probability densities. 
For our stellar evolution models in this study, we consider four continuous parameters:
\begin{equation}
\boldsymbol{\Theta} = \left\{ M,\alpha_\textsc{mlt}, X_\mathrm{i}, \tau \right\}.
\end{equation}
Here, $M$ denotes the stellar mass, $\alpha_\textsc{mlt}$ denotes the mixing length parameter, $X_\mathrm{i}$ denotes the initial mass fraction of hydrogen on the zero age main sequence (ZAMS), and $\tau$ denotes the stellar age.

The degree, to which we believe in our proposition, is the so-called posterior probability density, $p(\boldsymbol{\Theta}|\boldsymbol{D},\boldsymbol{I})$. It can be computed, using Bayes' theorem
\begin{equation}
p(\boldsymbol{\Theta}|\boldsymbol{D},\boldsymbol{I}) \propto p(\boldsymbol{\Theta}|\boldsymbol{I}) p(\boldsymbol{D}|\boldsymbol{\Theta},\boldsymbol{I}).
\end{equation}
Thus, $p(\boldsymbol{\Theta}|\boldsymbol{D},\boldsymbol{I})$ depends on the likelihood, $p(\boldsymbol{D}|\boldsymbol{\Theta},\boldsymbol{I})$, i.e. the degree, to which we believe in data given a certain set of model parameters. The likelihood will be addressed in Section~\ref{sec:likelihood}.
Furthermore, $p(\boldsymbol{\Theta}|\boldsymbol{D},\boldsymbol{I})$ depends on $p(\boldsymbol{\Theta}|\boldsymbol{I})$, i.e. our belief in the parameter values given our prior information. We will discuss our priors in detail in Section~\ref{sec:priors}.  

For any subset (${\boldsymbol{\Theta}_{A}}$) of the model parameters, we can obtain the posterior probability distribution function (PDF) by marginalization, i.e. integration over the remaining parameters ($\boldsymbol{\Theta}_{B}$):
\begin{equation}
p({\boldsymbol{\Theta}_{A}}|\boldsymbol{D},\boldsymbol{I}) = \int p(\boldsymbol{\Theta}|\boldsymbol{D},\boldsymbol{I}) \mathrm{d}{\boldsymbol{\Theta}_{B}},
\end{equation}
i.e. the PDFs for the investigated stellar parameters can be obtained by suitable projections in the parameter space. Here, $\boldsymbol{{\Theta}_{B}}$ is referred to as a nuisance parameter.

In order to map the posterior probability, we employ a Markov Chain Monte Carlo (MCMC) approach, performing a pseudo-random walk through the parameter space. The goal is to draw samples from the parameter space, in such a way that the resulting Markov chain converges towards the posterior probability distribution. 

Suppose that the previous entry in our Markov chain is $\boldsymbol{\mathcal{X}}^t$. The algorithm now randomly proposes a new point $\boldsymbol{\mathcal{Y}}$ in the parameter space
and computes a stellar evolution model, using this proposed set of parameters. Specifically, we compute the evolution of a star with a certain mass, mixing length parameter, and composition
until a given age is reached --- as any parameter, the age is drawn from a proposal distribution based on $\boldsymbol{\mathcal{X}}^t$.
The structure model that is obtained at the end of this evolution is the model that we want to compare to data:
based on the obtained stellar structure model and on observations, the MCMC algorithm calculates the likelihood for $\boldsymbol{\mathcal{Y}}$ and combines this with our prior information to obtain the associated posterior probability, $p(\boldsymbol{\mathcal{Y}}|\boldsymbol{D},\boldsymbol{I})$.
The algorithm then compares $p(\boldsymbol{\mathcal{Y}}|\boldsymbol{D},\boldsymbol{I})$ with $p(\boldsymbol{\mathcal{X}}^t|\boldsymbol{D},\boldsymbol{I})$, and accepts $\boldsymbol{\mathcal{Y}}$
as the next entry ($\boldsymbol{\mathcal{X}}^{t+1}$) in the Markov chain with a probability that is determined by this comparison.
An expression for the acceptance probability, i.e. the probability for accepting $\boldsymbol{\mathcal{Y}}$ as the next sample in the chain, is specified below. 
For now, it is sufficient to note that it is largely determined by the ratio of $p(\boldsymbol{\mathcal{Y}}|\boldsymbol{D},\boldsymbol{I})$ to $p(\boldsymbol{\mathcal{X}}^t|\boldsymbol{D},\boldsymbol{I})$.
If $\boldsymbol{\mathcal{Y}}$ is accepted, $\boldsymbol{\mathcal{X}}^{t+1}=\boldsymbol{\mathcal{Y}}$.
If $\boldsymbol{\mathcal{Y}}$ is rejected, $\boldsymbol{\mathcal{X}}^{t+1}$ is set equal to $\boldsymbol{\mathcal{X}}^t$. Having established the next entry of the Markov chain, the steps described above are repeated
by randomly proposing a new position in the parameter space.

In this paper, we employ the MCMC ensemble sampler published by \cite{emcee} --- this is an implementation of the procedure suggested by \cite{Goodman2010}.
Here, we include a short summary of some main aspects, referring to the cited papers for further details. 

Following the approach by \cite{Goodman2010}, we evolve an ensemble of $K$ walkers in parallel --- that is, we simultaneously perform $K$ coupled random walks through the parameter space.
The proposed distribution for any of these walkers is determined by the position of the $K-1$ other walkers.
{\color{black}The proposed next entry of the Markov chain for the $k$th walker at $\boldsymbol{\mathcal{X}}_k^t$ is $\boldsymbol{\mathcal{Y}}_k=\boldsymbol{\mathcal{X}}_w^t+\boldsymbol{\mathcal{Z}}$,
where $\boldsymbol{\mathcal{X}}_w^t$ denotes the current position of one of the $K-1$ other walkers ($w\neq k$), which has been randomly selected.} $\boldsymbol{\mathcal{Z}}$ is randomly drawn from a proposal
distribution, $g(z)$. In other words, $\boldsymbol{\mathcal{Z}}$ is a realization of $z$: 
\begin{equation}
g(z) = 
\begin{cases}
\frac{1}{\sqrt[]{z}} \qquad & \mathrm{for \: z \in \left[\frac{1}{a}, a \right]} \\
0 & \mathrm{else}.
\end{cases}
\end{equation}
{Here, $a$ is a tuneable (scale) hyperparameter, for which a value of 2 seems to ensure wide applicability \citep{Goodman2010}}.
The acceptance probability, i.e. the probability that the proposed next entry $\boldsymbol{\mathcal{Y}}_k$ in the Markov chain is accepted, is
\begin{equation}
\mathcal{A} = \min\left( 1 , \mathcal{Z}^{N-1} \frac{p(\boldsymbol{\mathcal{Y}}_k|\boldsymbol{D},\boldsymbol{I})}{p(\boldsymbol{\mathcal{X}}_k^t|\boldsymbol{D},\boldsymbol{I})} \right).
\end{equation}
{\color{black}Here, $N$ is the number of dimensions of the explored parameter space, i.e., in our case, $N=4$.
Having established the acceptance probability, one can then decide on rejecting or accepting $\boldsymbol{\mathcal{Y}}_k$ by drawing a random number ($\mathcal{R}$)
between zero and one from a uniform distribution.
If $\mathcal{A} \geq \mathcal{R}$, $\boldsymbol{\mathcal{Y}}_k$ is accepted. Otherwise, $\boldsymbol{\mathcal{Y}}_k$ is rejected.}


\subsection{Defining a likelihood for stellar modelling} \label{sec:likelihood}
The determination of a likelihood for stellar modelling has previously been discussed by \citet{Bazot2012} in their analysis of Alpha Centauri A. 
Due to their computational expense, MCMC methods are rarely used in stellar modelling. However, those authors have demonstrated that it is a robust and tractable strategy for selected targets. 
For completeness, we briefly summarize their arguments in the derivation of our likelihood function. 
We refer the reader to their work for a deep discussion on the philosophy of  MCMC  methods in the context of asteroseismic inference.

We possess no closed-form analytic formula that straightforwardly links the stellar model parameters to their observable quantities.
While there exist machine learning models and algorithms for rapid stellar evolution, they introduce additional error in their fitting, and they are not calibrated to our modelling choices (microphysics, boundary conditions, etc.,).
Here we wish to explore the impact of coupling 3D envelopes to 1D stellar models.
The necessary function evaluation and subsequent likelihood determination, therefore, takes of the order {\color{black}of} minutes to hours and requires solving the forward structure and oscillation equations.

{\color{black}For the purpose of this paper, a stellar evolution code is a function, $f(\boldsymbol{\Theta})$, that maps $\boldsymbol{\Theta}$ to their predicted observable quantities ($\boldsymbol{\Tilde{D}}$).
Specifically, we consider the mapping of $\boldsymbol{\Theta}$ to a set of seismic and spectroscopic properties:}
\begin{equation}
    \boldsymbol{\Tilde{D}} = \left\{ \Teff, \Dnu,  \left< r_{01} \right>, \rm{etc.,}\right\}. 
\end{equation}
However, as is the case when comparing theoretical calculations ($\boldsymbol{\Tilde{D}}$) with observable data ($\boldsymbol{D}$),  there is uncertainty in our measurements.
We, therefore, have the situation 
\begin{equation}
     f(\boldsymbol{\Theta})= \boldsymbol{\Tilde{D}} \qquad \mid \boldsymbol{D}=\boldsymbol{\Tilde{D}}+\boldsymbol{\epsilon} 
\end{equation}
where $\boldsymbol{\epsilon}$ is due to measurement error or deficiencies in the model.
For asteroseismic inference, we seek to invert the problem, i.e.
\begin{equation}
    \boldsymbol{D} = f(\boldsymbol{\Theta}) +\boldsymbol{\epsilon}, 
\end{equation}
which is typically found by comparing models to the data and minimizing an objective function such as $\chi^2$. 
In the case of MCMC, we aim to recover the underlying distribution rather than to optimize.  
For our purposes, we can interpret $\boldsymbol{\epsilon}$  as a realization of a random process. 

We seek to define a likelihood, $\mathcal{L}=p(\boldsymbol{D}|\boldsymbol{\Theta}\boldsymbol{I})$, which is a function of the data, $\boldsymbol{D}$, given the a model parameterized by $\boldsymbol{\Theta}$. {\color{black}We assume that the measurement ($D_j$) of any of the $J$ spectroscopic quantities is uncorrelated with the remaining measurements and that the associated noise is Gaussian.
We can hence write the contribution of the spectroscopic constraints to the likelihood, $\mathcal{L}=p(\boldsymbol{D}|\boldsymbol{\Theta},\boldsymbol{I})$, as
\begin{equation}
\mathcal{L}_{\rm{spec}} =  \prod_{j=1}^{J} \frac{1}{\sqrt{2\upi \sigma_j^2}} \times \exp \left( - \frac12 \left[ \frac{f(\boldsymbol{\Theta})_j-D_j}{ \sigma_j} \right]^2 \right).
{\label{eq:likeli1}}
\end{equation}
Here, the sum runs over all spectroscopic quantities, and $\sigma_j$ denotes the corresponding standard deviation for the observation $D_j$.
The analysis presented in this paper involves a single spectroscopic quantity, $T_\mathrm{eff}$.
As mentioned above, we ignore the diffusion of elements and consequently assume the metallicity to be fixed, throughout the evolution. Thus, all models do, by default have the same $\mathrm{[Fe/H]}$, which means that an implementation of
$\mathrm{[Fe/H]}$ into the likelihood would not alter the obtained PDFs.

While there are other observational constraints, including spectroscopic measurements of $L$ and $\log g$, that we could make use of, they are not all independent pieces of information about the star.
Some quantities are already well constrained by the asteroseismic measurements, making further constraints redundant and potentially leading to overfitting. {\color{black}Furthermore, while interferometry allows for the precise determination of the luminosity of Alpha Centauri A and B, this is not generally the case for all seismic objects. Similarly, while $\log g$ is well known for these stars, it is often the case that the spectroscopically and asteroseismically determined $\log g$ differ. As a result, spectroscopic constraints on $\log g$ are often trained or calibrated based on seismology \citep{2013A&A...556A..59H} which does not result in a new independent measurement. Excluding $L$ and $\log g$ from our likelihood thus allows for a differential study of a well-known target without loss of generality.}

For the $Q$ asteroseismic parameters, we generalise the univariate normal distribution to higher dimensions, $\boldsymbol{D} \sim \mathcal{N}_p(f({\boldsymbol{\Theta}}),\mathbfss{C})$,
to take their co-variances ($\mathbfss{C}$) into account:
\begin{eqnarray}
\hspace{-1.5cm}\mathcal{L}_{\rm{seis}} &=& \left(2\upi\right)^{-Q/2} \left|\mathbfss{C} \right|^{-Q/2}
\times \nonumber\\
&& \hspace{-1.cm} \exp \left( -  \frac12 \sum_{q=1}^{Q} \left[f(\boldsymbol{\Theta})_q-D_q\right]^T \mathbfss{C}^{-1} \left[f(\boldsymbol{\Theta})_q-D_q\right] \right).
{\label{eq:likeli2}}
\end{eqnarray}
Here, the sum runs over all seismic quantities, and $\sigma_q$ denotes the corresponding standard deviation for the observation $D_q$ (cf. Section~\ref{sec:astroprop}).
We combine $\mathcal{L}_{\rm{spec}}$ and $\mathcal{L}_{\rm{seis}}$, i.e. Eqs~(\ref{eq:likeli1}) and (\ref{eq:likeli2}), to determine a final likelihood for the model: $\mathcal{L}=\mathcal{L}_{\rm{spec}}\mathcal{L}_{\rm{seis}}$.}

The reason for the two-component likelihood is that the asteroseismic quantities are highly correlated and do not provide independent pieces of information about stellar properties \citep{2017ApJ...839..116A, 2018arXiv180807556R}. 
We, therefore, first determine the correlation matrix for every measured frequency table.  
Either the correlation matrix is read in from MCMC fitting to the power spectrum \citep{Handberg2011}, or it is determined in situ directly from Monte Carlo realizations of the measured frequencies.
In the case of the latter, we perturb each measured frequency independently with Gaussian noise according to its measurement uncertainty.
We typically perform 10,000 {realizations} from which we calculate asteroseismic separations and ratios.  
From these, the correlation matrix is determined for use in Eq.~(\ref{eq:likeli2}).

\subsection{Priors} \label{sec:priors}

Bayesian statistics allows us to include prior information in the posterior probability distribution.
If the prior PDF of each of the $N$ parameters $\Theta_n$, to which we will refer as $p(\Theta_n|\boldsymbol{I})$, are mutually independent
\begin{equation}
p(\boldsymbol{\Theta}|\boldsymbol{I}) = \prod_{n=1}^N p(\Theta_n|\boldsymbol{I}).
\end{equation}
In the present case, $\Theta_n$ is either $M$, $\alpha_\textsc{mlt}$, $X_\mathrm{i}$, or the stellar age.
For the stars investigated in this paper, we mostly impose uniform priors:
\begin{equation}
p(\Theta_n|\boldsymbol{I}) = 
\begin{cases}
\frac{1}{\Theta_n^\mathrm{max}-\Theta_n^\mathrm{min}} \qquad & \mathrm{for} \,\, \Theta_n^\mathrm{max}< \Theta_n <\Theta_n^\mathrm{max}, \\
0, & \mathrm{else}. \label{eq:uniform}
\end{cases}
\end{equation}
We note that Eq.~(\ref{eq:uniform}) constitutes an improper prior, as it is not normalized. We further impose a Gaussian prior on the mass of the star,
as we investigate stars that are situated in a binary, i.e. for which detailed measurements of the mass are available:
\begin{equation}
p(\Theta_n|\boldsymbol{I}) = \frac{1}{\sqrt{2\upi \sigma_n^2}} \exp \left( \frac{-(\theta_n-\mu_n)^2}{2\sigma_n^2} \right),
\end{equation}
where $\mu_n$ and $\sigma_n$ denote the mean and the standard deviation of the Gaussian prior, respectively.


\subsection{Specification of priors}

We demand that the age of the star may not exceed the age of the Universe, setting the upper boundary on a uniform prior of the age to $13.8\,$Gyr \cite[cf.][]{Planck2016}.

We set the lower boundary of the uniform prior for $\alpha_\mathrm{mlt}$ to 1.0. The upper boundary is set to 2.5 and 10.0 for the standard and coupled stellar models, respectively. In both cases, we have chosen the width of the uniform prior based on solar calibrations using the same method. For standard models, a solar calibration yields $\alpha_\mathrm{mlt}=1.78$ (cf. \citetalias{Joergensen2018}). Allowing $\alpha_\mathrm{mlt}$ to deviate strongly from this may lead to multimodal distributions, due to the degeneracy between $\alpha_\mathrm{mlt}$ and other parameters. This is the argument for using a lower upper boundary than in the case of the coupled stellar models.

In the case of the coupled stellar models, the unconventionally high upper boundary for the mixing length parameter is motivated by the results published by \citetalias{Joergensen2018}: a solar calibration that employs coupled stellar models yields a mixing length parameter of 3.30. {Physically, this seemingly high value of $\alpha_\textsc{mlt}$ can be explained in terms of entropy: $\alpha_\textsc{mlt}$ must bridge the entropy jump between the deep asymptotic adiabat and the outer boundary of the interior model. In the case of coupled models, the outer boundary is placed in the nearly-adiabatic layers. Thus, most of the entropy jump near the stellar surface takes place in the appended $\langle \mathrm{3D} \rangle$-envelope, and MLT is only employed in a narrow layer far below the photosphere. In standard stellar models, on the other hand, MLT is needed to recover the entire entropy jump.
The mixing length parameter must hence encompass very different information in the two scenarios. We come back to this issue in Section~\ref{sec:discussion}.}

As regards to the mass, we impose a uniform prior in combination with a Gaussian prior. The latter reflects the radial velocity measurements (orbital analysis) of the targets.
The former is motivated by the restrictions set by the Stagger grid with the upper limit chosen in order to avoid extrapolating from the simulations near the zero-age main sequence (ZAMS). 
Note that the metallicity of the star shifts the ZAMS, i.e. the upper limit on the mass decreases with metallicity. 
We set the lower and upper boundaries on the mass to $0.65\,\mathrm{M}_\odot$ and $1.30\,\mathrm{M}_{\odot}$, respectively. 
All boundaries of the employed uniform priors are summarized in Table~\ref{tab:uniform}.
 
\begin{table*}
    \centering
    \caption{Upper and lower boundary for the uniform priors.} 
    \label{tab:uniform}
    \begin{tabular}{lccccccccccccccccccccc} 
        \hline
        & $M$ [$\mathrm{M_\odot}$] & $\alpha_\textsc{mlt}$ & $X_\mathrm{i}$ &  Age [Myr]\\
        \hline
        Upper limit & 1.30 & 2.5 (1D), 10.0 (3D) & 0.80 & 13800 \\ [2pt]
        Lower limit & 0.65 & 1.0 & 0.60 & 50 \\ [2pt]
        \hline
\end{tabular}
\end{table*}

Based on \cite{Pourbaix2002}, \cite{Pourbaix2016} found the mass of Alpha Centauri A and B to be $1.133\pm0.0050 \, \mathrm{M}_\odot$ and $0.972\pm0.0045\,\mathrm{M}_\odot$, respectively. We include this information as a prior, truncating the Gaussians, using the upper and lower limit listed in Table~\ref{tab:uniform}.

The stellar evolution code terminates when it reaches the end of the red giant branch (RGB) or leaves the parameter space covered by the Stagger grid. Very high propositions for the stellar age may, therefore, force the stellar evolution code to abort the computation, before computing a structure for the desired combination of parameters.
To circumvent this problem, we attribute a posterior probability of zero to models, for which
\begin{equation}
\Delta \nu = \Delta \nu_\odot \left(\frac{M}{\mathrm{M}_\odot}\right)^{0.5} \left(\frac{T_\mathrm{eff}}{\mathrm{T}_{\mathrm{eff},\odot}}\right)^{3}\left(\frac{L}{\mathrm{L}_\odot}\right)^{-0.75} < 20\,\mu\mathrm{Hz}.
\end{equation}
Here $\Delta \nu$ is an approximation to $\Delta_0$ (cf. Eq.~\ref{eq:large}), {\color{black}$M_\odot$ is the mass of the present-day Sun, and $L/L_\odot$ denotes the luminosity in units of the solar luminosity.} In this paper, we assume that $T_{\mathrm{eff},\odot}=5777\,$K and that $\Delta \nu_\odot =135\,\mu$Hz.


\subsection{Initial conditions, convergence and burn-in}

For each star, we started all walkers with initial parameters that were close to the expected maximum of the posterior probability distribution. This approach is suggested by \cite{emcee} since there is the risk that the walkers get stuck in a local maximum when sampling multi-modal probability distributions. 

The initial parameters were drawn from normal distributions, for which the standard deviation corresponded to 2 per cent of the mean value. 
The mean of $X_\mathrm{i}$, $\alpha_\mathrm{mlt}$, and the age for the initial conditions of the walkers were consistently set to $0.7$, $1.8$ and $5.0\,$Gyrs, respectively.

To ensure that the obtained posterior probability distributions do not reflect the initial conditions of the walkers, we allow for a burn-in phase.  
We have estimated the integrated autocorrelation time that describes the number of samples that is required for the walkers to forget about their initial conditions \citep{emcee}.
Due to the relatively low number of samples obtained per walker, we have estimated this quantity, using an autoregressive-moving-average {\color{black}(ARMA) model. ARMA models aim to describe a time series, i.e. the Markov chain,
by assuming that each element of the time series depends linearly on previous entries as well as on the current and previous error terms.
In other words, ARMA models are linear approximations that provide a simplified description of stochastic processes.
Following this approach,} we determined the autocorrelation time to be less than roughly 200 samples per walker. 
We, nevertheless, adopt a more conservative approach and allow a burn-in phase of 250 models per walker. In all cases, our exploration of the parameter space employs 16 walkers, which means that we discard 4000 samples for each run.


\section{Stellar parameter estimates}

For the determination of the stellar parameters, we do not vary $\mathrm{[Fe/H]}$ but keep it fixed throughout the evolution.
According to \cite{Thevenin2002}, $\mathrm{[Fe/H]}$ for Alpha Centauri A and B are $0.20\pm0.02$ and $0.23\pm 0.03$, respectively. Furthermore,
\cite{Nsamba2018b} estimate $\mathrm{[Fe/H]}$ of Alpha Centauri A to be $0.23\pm0.05$. These measurements are hence consistent with the hypothesis that both stars have the same composition. We, therefore, set $\mathrm{[Fe/H]}=0.25$ for both the standard and the coupled models, based on the quoted spectroscopic constraints. 

As regards the coupled stellar models, \textsc{garstec} performs the interpolation in $T_\mathrm{eff}$ and $\log g$ at each time-step based on 29 interpolated $\langle \mathrm{3D} \rangle$-envelopes with $\mathrm{[Fe/H]}=0.25$. These interpolated envelopes have been evaluated, using the interpolation in composition proposed by \cite{Joergensen2019}. This paper hence presents the first analysis of coupled models at non-solar metallicity.

We have computed the posterior mean and use the posterior standard deviations as the associated credible intervals {\color{black}when listing parameter estimates}.


\subsection{Alpha Centauri A} \label{sec:ACA}

For the primary star in the binary, we used the observed frequencies ($\ell=0-2$) by \cite{Meulenaer2010} and spectroscopic constraints by \cite{Kervella2017} {\color{black}to determine the likelihood}: we set $T_\mathrm{eff}=5795\pm 57\,$K, using three times the statistical uncertainty quoted by \cite{Kervella2017} to allow for systematic uncertainties. This effective temperature is in good agreement with that found by other authors \citep[e.g.][]{Thevenin2002}.
We set $\nu_\mathrm{max} = 2300\pm50\,\mu$Hz, which can be translated into a mass of $1.11\,\mathrm{M}_\odot$, based on asteroseismic scaling relations. We use this mass estimate for the mean of the Gaussian mass distribution, from which the initial mass estimates of the walkers are drawn.

\begin{figure*}
\centering
\includegraphics[width=0.85\linewidth]{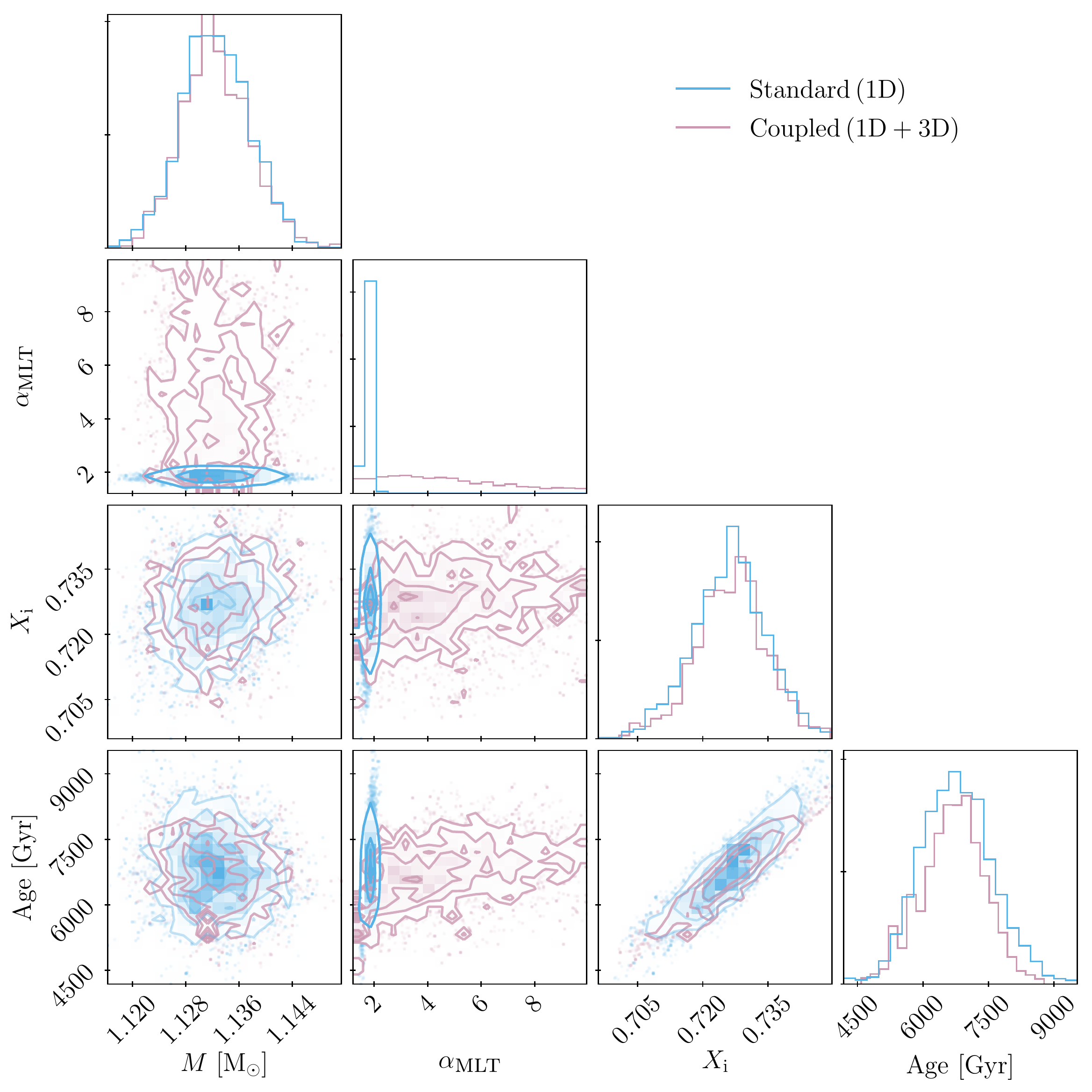}
\caption{Corner plot showing different projections of the samples and of the posterior probability distributions for the stellar parameters ($M$, $\alpha_\textsc{mlt}$, $X$, and age) of Alpha Centauri A. The plot is based on 7200 standard stellar models (blue) and 6848 coupled stellar models (purple).
}
\label{fig:cornerACA}
\end{figure*}

After excluding a burn-in phase, we obtained 7200 and 6848 samples for the standard and coupled models, respectively. The posterior probability distributions are summarized in Table~\ref{tab:posterior} as well as in the corner plot shown in Fig.~\ref{fig:cornerACA}, i.e. the marginalized probability densities.

\begin{table*}
    \centering
    \caption{Summary of the posterior probability distributions found by using standard (1D) and coupled (3D) stellar models, respectively. We list the median of the corresponding probability distributions as the parameter estimate. The associated errors correspond to the 16th and 84th percentile.}
    \label{tab:posterior}
    \begin{tabular}{lccccccccccccccccccccc} 
        \hline
         & $M^{\mathrm{1D}}$ [$\mathrm{M}_\odot$] &$M^{\mathrm{3D}}$ [$\mathrm{M}_\odot$] & $\alpha^\mathrm{1D}_\textsc{mlt}$ & $\alpha^\mathrm{3D}_\textsc{mlt}$ & $X^\mathrm{1D}_\mathrm{i}$ & $X^\mathrm{3D}_\mathrm{i}$ & Age, 1D [Gyr] & Age, 3D [Gyr] \\
        \hline
        $\alpha$ Cen A  & $1.132^{+0.005}_{-0.005}$ & $1.132^{+0.006}_{-0.004}$ & $1.78^{+0.10}_{-0.11}$ & $4.30^{+2.87}_{-1.99}$ & $0.726^{+ 0.008}_{-0.008}$ & $0.727^{+ 0.008}_{-0.008}$ & $6.76^{+0.80}_{-0.80}$ & $6.74^{+0.69}_{-0.80}$ \\ [5pt]
        $\alpha$ Cen B  & $0.971^{+0.005}_{-0.004}$ & $0.972^{+0.004}_{-0.005}$ & $1.95^{+0.10}_{-0.10}$ & $6.41^{+2.37}_{-2.60}$ & $0.733^{+ 0.007}_{-0.008}$ & $0.731^{+ 0.008}_{-0.008}$ & $5.37^{+0.93}_{-0.91}$ & $4.86^{+0.77}_{-0.75}$ \\ [5pt]
        \hline
\end{tabular}
\end{table*}

We find that the use of standard and coupled stellar models lead to parameter estimates for $M$, $X_\mathrm{i}$, and the stellar age that are mutually consistent. As regards $\alpha_\textsc{mlt}$, the standard model yields a value of $1.78\pm0.11$, which is in good agreement with standard solar calibrations. As we ignore some input physics that enter a proper solar calibration, including metal diffusion, any deviation from the solar calibration value may reflect this neglect as well as variations in the mixing length with the stellar global parameters \citep[e.g.][]{Tayar2017,Trampedach2014b}.

The coupled models result in an estimate for $\alpha_\textsc{mlt}$ of $4.30^{+2.87}_{-1.99}$. This is consistent with findings of \citetalias{Joergensen2018}: in the coupled models, MLT is only used to describe a thin nearly adiabatic layer leading to higher values of $\alpha_\textsc{mlt}$, as argued above. The large error bars show that actual value of $\alpha_\textsc{mlt}$ becomes less relevant. We refer to Section~\ref{sec:discussion} for a detailed discussion thereof.

\begin{figure}
\centering
\includegraphics[width=0.85\linewidth]{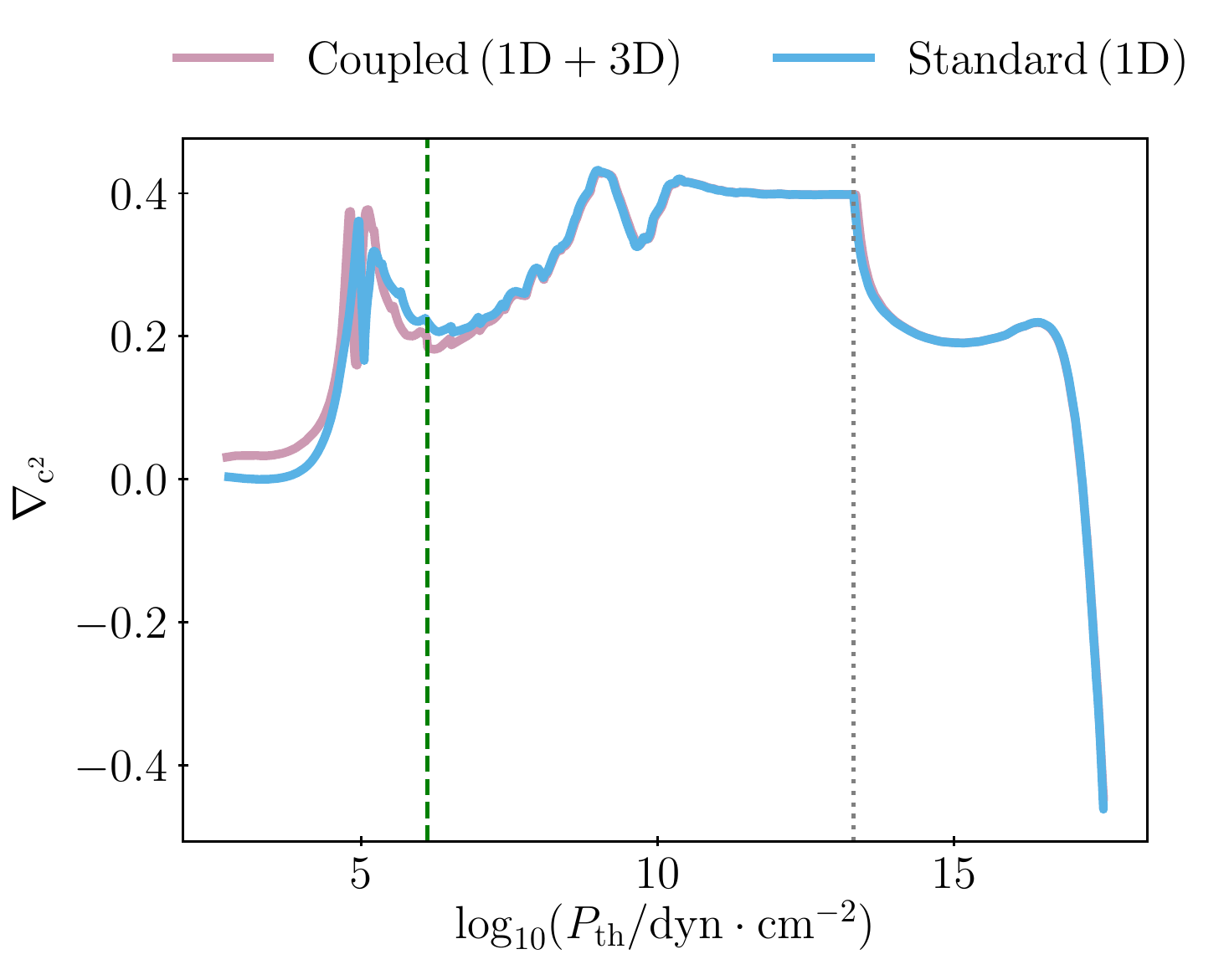}
\caption{The gradient of the squared sound speed as a function of the thermal pressure for Alpha Centauri A. The figure includes the best-fitting standard (solid blue line) and coupled (solid purple line) model. The dotted grey line shows the location of the base of the convective envelope. The dashed green line indicates the position of the matching point in the coupled model.
}
\label{fig:c2_ACA}
\end{figure}

As for the Sun, the use of 3D envelopes does not significantly alter the depth of the convective envelope \citepalias[see table~1 in ][]{Joergensen2018}: we find that the energy transport becomes radiative, i.e. that the Schwarzschild criterion for convective instability ($\nabla_\mathrm{rad}>\nabla_\mathrm{ad}$) is no longer fulfilled, for radii below 0.726$\,R$ and 0.722$\,R$ for the best-fitting standard and coupled models, respectively. Here, $R$ denotes the photospheric radius of the star. We illustrate this in Fig.~\ref{fig:c2_ACA} by plotting \citep{jcd1991,jcd2011}
\begin{equation}
\nabla_{c^2} = \frac{\mathrm{d} \ln c^2}{\mathrm{d} \ln P_\mathrm{th}} \approx \nabla - \frac{\mathrm{d}\ln \mu}{\mathrm{d} \ln P_\mathrm{th}}
\end{equation}
of the best-fitting standard and coupled models, i.e. the models with the highest posterior probability, as a function of the thermal pressure ($P_\mathrm{th}$). Here, $c$ and $\mu$ denote the sound speed and the mean molecular weight, respectively.

The abrupt transition in $\nabla_\mathrm{c^2}$ near $0.7\,R$ indicates the base of the convective envelope. As can be seen from the figure, the interior structures of the best-fitting models are rather similar. Only the outer convective layers strongly differ.
The deeper interior of the models, on the other hand, are not affected.
This is, again, in very good agreement with the result found in the case of the present-day Sun. 

{\color{black}Other quantities, such as the temperature stratification of the best-fitting standard and coupled model, likewise} converge in the deep interior below the matching point. Despite very different values of $\alpha_\textsc{mlt}$, the two models hence reach the same deep asymptotic adiabat. We address this issue further in Section~\ref{sec:discussion}.

For the best-fitting standard stellar model, $M= 1.133 \,\mathrm{M}_\odot$, $\alpha_\textsc{mlt}=1.79$, $X_\mathrm{i}=0.727$ ($Y_\mathrm{i}=0.249$), and the age is $6.85\,$Gyr. For the best-fitting coupled stellar model, $M= 1.131 \,\mathrm{M}_\odot$, $\alpha_\textsc{mlt}=5.19$, $X_\mathrm{i}=0.729$ ($Y_\mathrm{i}=0.248$), and the age is $7.02\,$Gyr.
Note that these parameter estimates correspond to the values that result in the highest posterior probability, whereas the values presented in Table~\ref{tab:posterior} denote the medians of the corresponding probability distribution functions.

All in all, for MS stars, such as the Sun or Alpha Centauri A, the use of 3D envelopes does not seem to lead to parameter estimates that deviate significantly from those obtained using MLT with the exception of the mixing length parameter.

\begin{figure}
\centering
\includegraphics[width=0.85\linewidth]{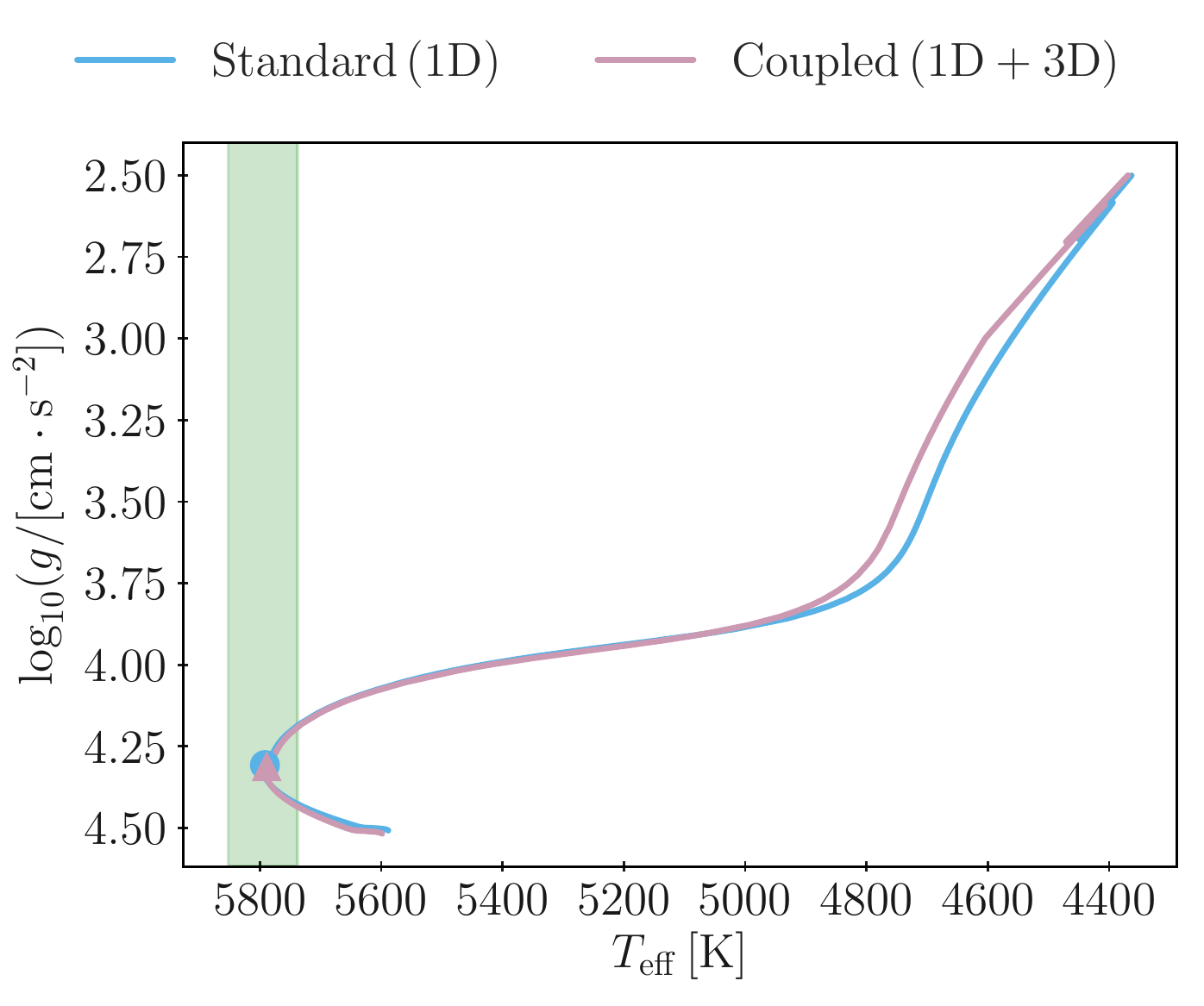}
\caption{Predicted stellar evolution of Alpha Centauri A, for the best-fitting standard (blue solid line) and coupled (purple dashed line) stellar models. The shaded green area shows the $68\,\%$ credible interval of the observational constraint on the effective temperature. The global parameters of the best-fitting standard (blue circle) and coupled (purple triangle) structure models are likewise included.
}
\label{fig:Ahrd_1D3D}
\end{figure}

{\color{black}However,} as discussed by \citetalias{JoergensenWeiss2019}, the coupling of 1D and 3D simulations affects the subsequent evolution. In the case of the Sun, it shifts the predicted turn-off point, i.e. the end of the MS in the HR diagram. In addition, the effective temperature on the RGB is altered. While the stellar parameter estimates of MS stars are mostly unaffected by the change in the outer boundary conditions, it is to be expected that the coupling of 1D and 3D simulations has a larger effect for the parameter estimates of later evolutionary stages. 
For Alpha Centauri A, the changes in the evolution track are consistent with this picture on the RGB. However, the best-fitting model lies close to the turn-off point, for which no shift is seen. The largest discrepancy in $T_\mathrm{eff}$ at constant $\log(g)$ is roughly $40\,$K for $\log(g)>2.5$. This is shown in Fig.~\ref{fig:Ahrd_1D3D}. 

Regarding Fig.~\ref{fig:Ahrd_1D3D}, we note that the evolution tracks of the coupled models show kinks --- predominantly on the RGB. This issue is discussed in detail in the follow-up paper of \citetalias{Joergensen2018}. We attribute these kinks to interpolation errors that reflect the low sampling of the relevant parameter space and calls for a refinement of the Stagger grid. 

\begin{figure}
\centering
\includegraphics[width=0.85\linewidth]{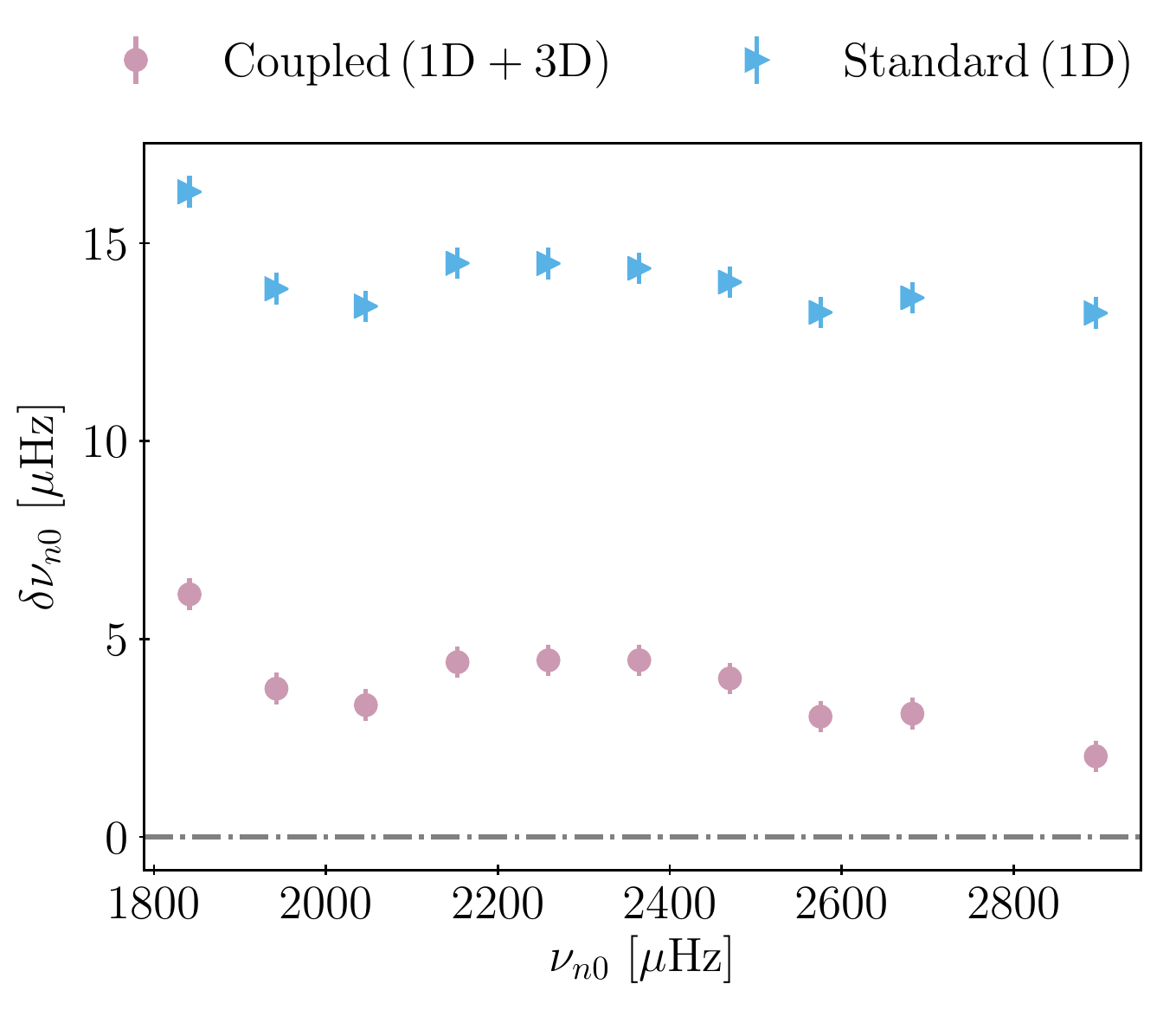}
\caption{Frequency difference between observations and the best-fitting models for Alpha Centauri A. We have only included radial modes ($\ell=0$).
}
\label{fig:Freq_ACA}
\end{figure}

Furthermore, we note that the p-mode frequencies of the best-fitting models are systematically lower than the observed frequencies. This is illustrated in Fig.~\ref{fig:Freq_ACA}. In the case of the standard models, this is slightly at odds with the expected behaviour: due to the surface effect, the model frequencies are expected to be systematically too high rather than too low. Also, for the standard stellar models, we would expect the frequency offset to increase with increasing frequency. This discrepancy reflects the fact that the individual frequencies neither enter the likelihood nor the priors and that they are hence not required to be reproduced. Only the frequency ratios and separations must be recovered. The latter requirement does not necessitate {\color{black}but rather averts the recovery of a frequency-dependent surface effect, as discussed in Section~\ref{sec:astroprop}:}
as can be seen from Fig.~\ref{fig:Freq_ACA}, the systematic offset is constant as a function of frequency. 
In addition, the frequency ratios enter our likelihood, and it is well-known that an evaluation of model parameters based on frequency ratios lead to different results than when using individual frequencies \citep[e.g.][]{Aguirre2013,Basu2018,Nsamba2018}.
Of course, we do also not expect our best-fitting model to match observations perfectly. After all, it is just the best model in the corresponding Markov chain given the restricted input physics and our selection criteria.
The inability of the model to recover the correct individual frequencies is hence a trade-off that we have found acceptable for the purposes of this exercise.
Moreover, in this paper, we merely present a differential study whereby we compare models that employ two physically different descriptions of the outer boundary layers. The likelihood was, therefore, chosen in order to facilitate a one-to-one comparison of the respective predictions. 
The fact that the best-fitting models do not match all observations of the target stars is, therefore, an acceptable {\color{black}drawback}.

As regards the burn-in phase, we obtain parameter estimates that are consistent with those in Table~\ref{tab:posterior}, if we exclude further samples. 
Finally, as regards the obtained mass estimates, we note that we achieve statistical agreement with the dynamic masses from the binary orbit. However, we note that the likelihood is seemingly not sufficiently informative to put further constraints on the mass beyond those imposed by radial velocity measurements. In other words, the posterior probability distribution of the mass is dominated by the prior. This does, however, not affect the main conclusions drawn from the analysis presented in this paper: we address the influence of the mixing length parameter when coupling 1D and 3D models.


\subsection{Alpha Centauri B} \label{sec:ACB}

For the secondary star, we used the observed frequencies ($\ell=0-2$) by \cite{Kjeldsen2005} in combination with the spectroscopic constraints by \cite{Kervella2017} {\color{black}to determine the likelihood}: we hence set $T_\mathrm{eff}=5231\pm 63\,$K, once again adopting $3\sigma$ uncertainties. We set the mean of the initial Gaussian mass distribution to $0.96\,\mathrm{M}_\odot$. 

\begin{figure*}
\centering
\includegraphics[width=0.85\linewidth]{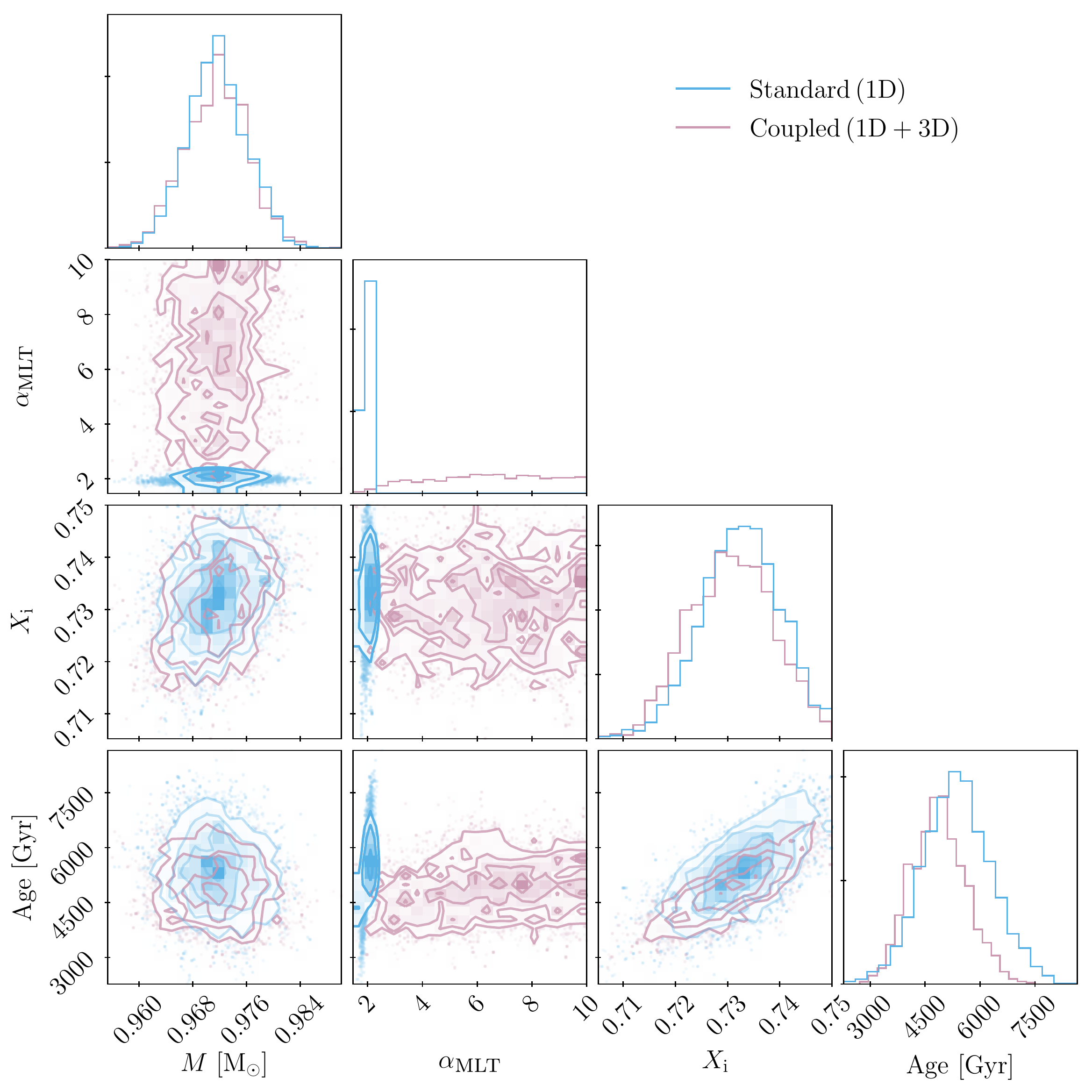}
\caption{Same as Fig.~\ref{fig:cornerACA} but for Alpha Centauri B. The plot is based on 7200 standard stellar models (blue) and 6784 coupled stellar models (purple).
}
\label{fig:cornerACB}
\end{figure*}

After excluding a burn-in phase, we obtained 7200 and 6784 samples for the standard and coupled models, respectively. The posterior probability distributions are summarized in Table~\ref{tab:posterior} alongside those of Alpha Centauri A as well as in the corner plot shown in Fig.~\ref{fig:cornerACB}.

From the obtained posterior probability distributions, we draw the same qualitative conclusions as for Alpha Centauri A: the use of coupled and standard stellar models lead to consistent parameter estimates with the exception of the mixing length parameter: the posterior probability of the coupled models are found to be rather insensitive to $\alpha_\textsc{mlt}$.

\begin{figure}
\centering
\includegraphics[width=0.85\linewidth]{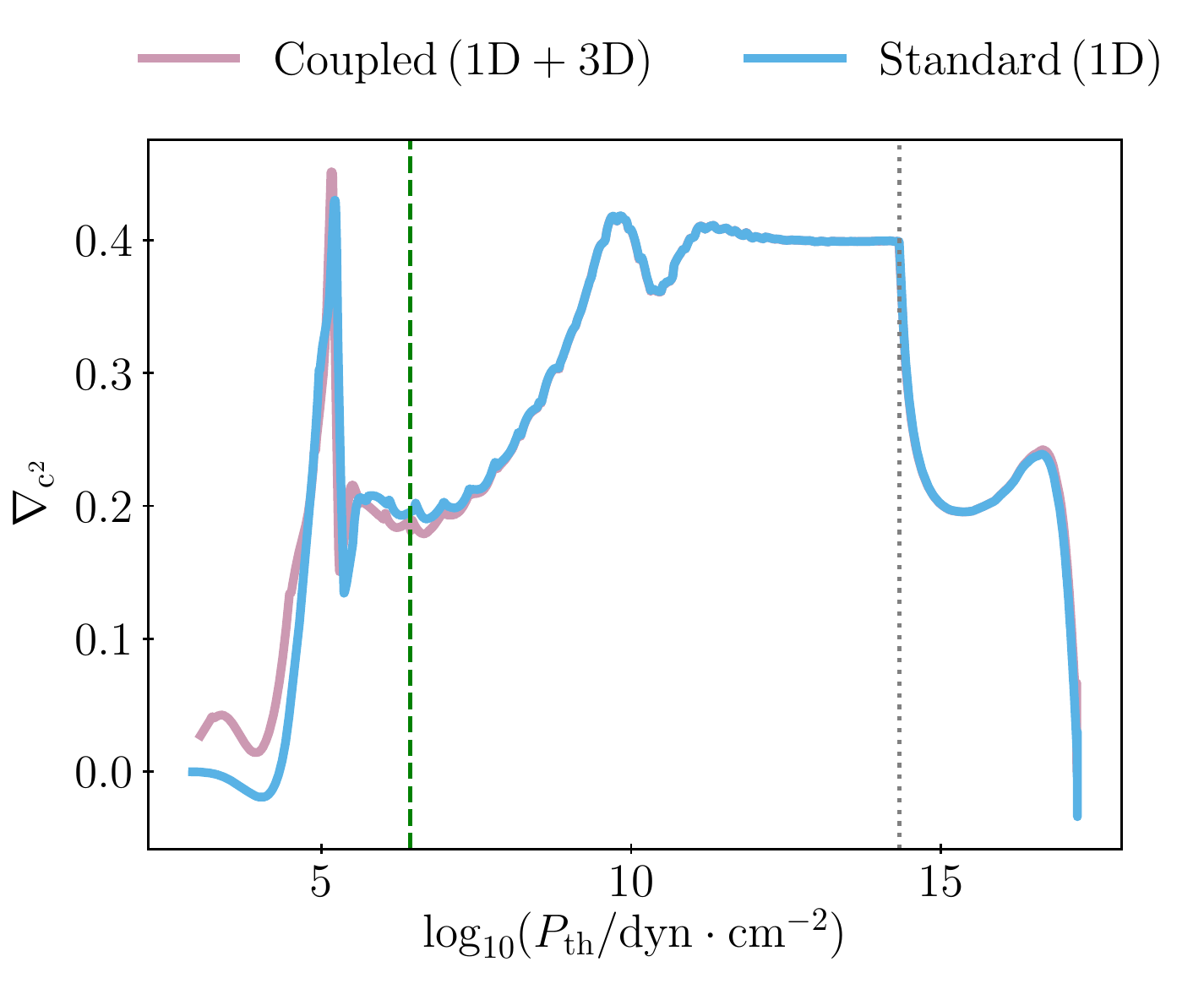}
\caption{Same as Fig.~\ref{fig:c2_ACA} but for Alpha Centauri B.
}
\label{fig:c2_ACB}
\end{figure}

\begin{figure}
\centering
\includegraphics[width=0.85\linewidth]{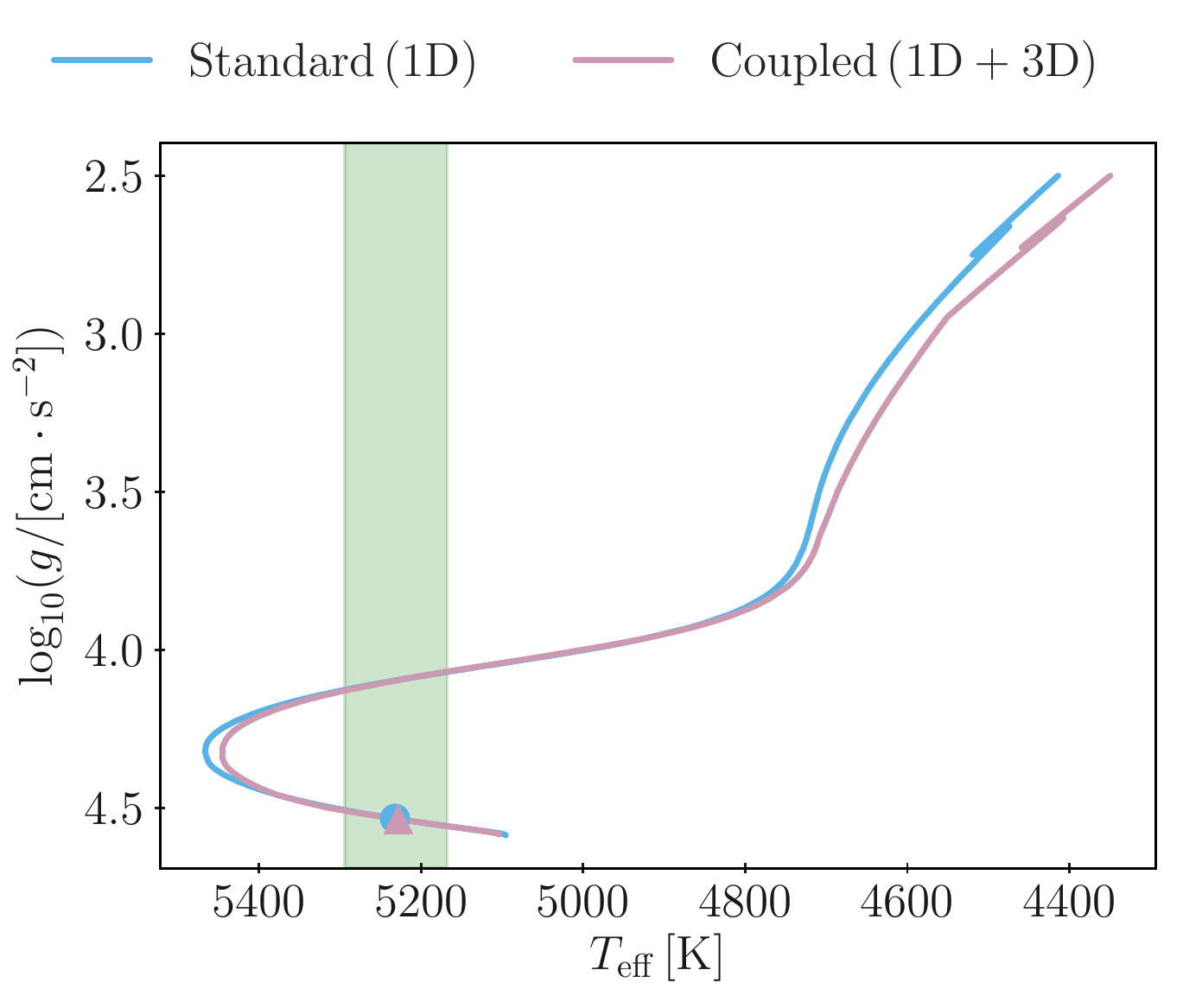}
\caption{Same as Fig.~\ref{fig:Ahrd_1D3D} but for Alpha Centauri B.
}
\label{fig:hrd_1D3D}
\end{figure}

\begin{figure}
\centering
\includegraphics[width=0.85\linewidth]{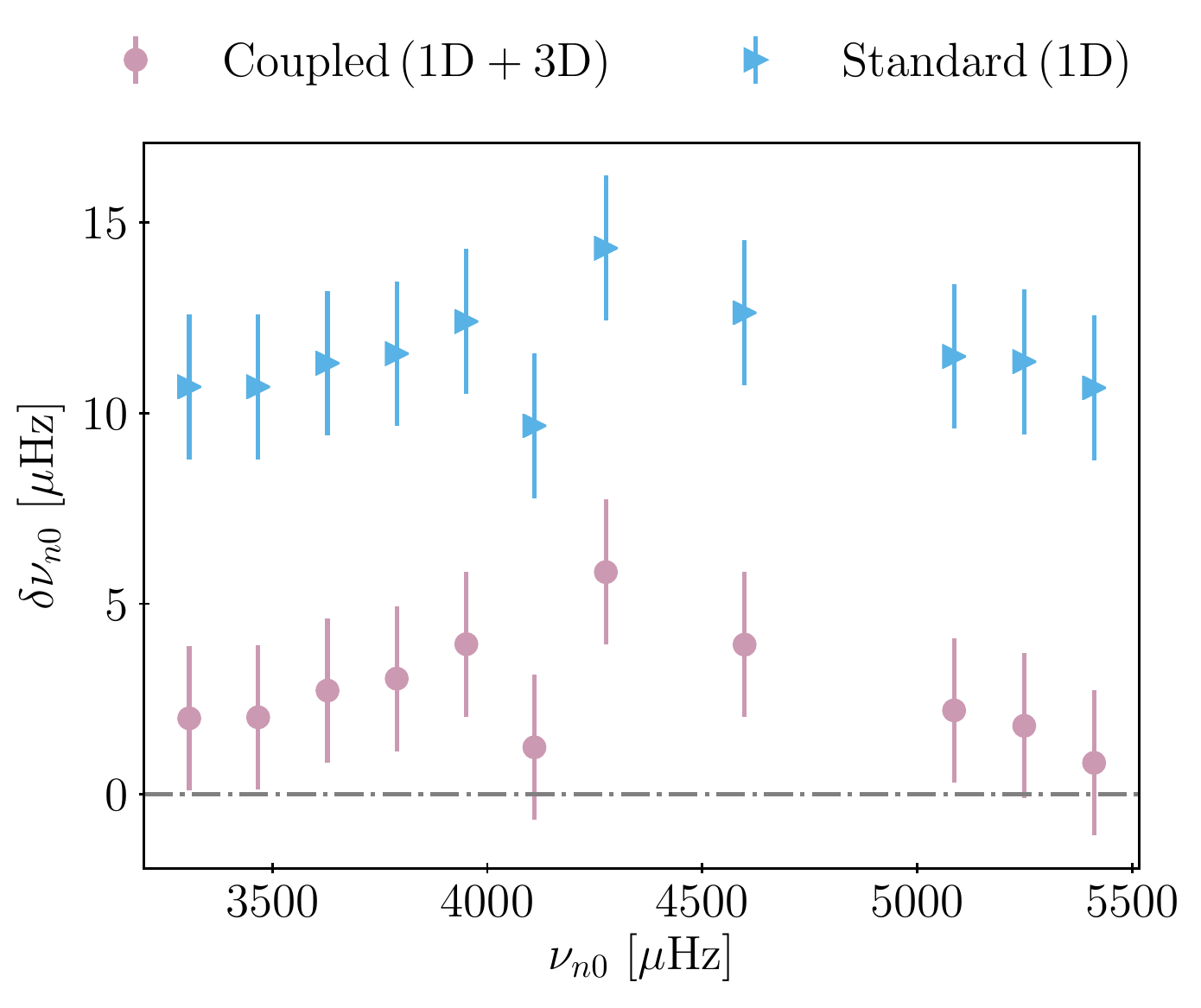}
\caption{Same as Fig.~\ref{fig:Freq_ACA} but for Alpha Centauri B.
}
\label{fig:Freq_ACB}
\end{figure}

Furthermore, the best-fitting models for Alpha Centauri B behave in the same way as for Alpha Centauri A: the interior structure of the best-fitting coupled and standard models are rather similar; only the outer layers deviate significantly (cf. Fig.~\ref{fig:c2_ACB}). While the best-fitting models are hence rather similar, the later evolution differs more significantly (cf. Fig.~\ref{fig:hrd_1D3D}). In contrast to the case of Alpha Centauri A, and in accordance with the solar case, we see that the use of coupled models shifts the turn-off point. Finally, the reconstruction of the individual frequencies is acceptable, considering that these did not enter the likelihood (cf.~Fig.~\ref{fig:Freq_ACB}) as argued above. {\color{black}As in the case of Alpha Centauri A, we would thus like to underline that we do, by no means, claim that our best-fitting model gives a complete description of Alpha Centauri B.  The best-fitting model is merely the best model within the Markov chain based on the defined likelihood and priors.}

For the best-fitting standard stellar model, $M= 0.972 \,\mathrm{M}_\odot$, $\alpha_\textsc{mlt}=1.96$, $X_\mathrm{i}=0.731$ ($Y_\mathrm{i}=0.245$), and the age is $5.39\,$Gyr. For the best-fitting coupled stellar model, $M= 0.972 \,\mathrm{M}_\odot$, $\alpha_\textsc{mlt}=7.88$, $X_\mathrm{i}=0.729$ ($Y_\mathrm{i}=0.247$), and the age is $4.89\,$Gyr. The base of the convection zone is placed at $0.6855\,$R and $0.6875$ for the best-fitting standard and coupled model, respectively.

We note that the ages of the primary and secondary star in the binary are comparable but not identical. This is a recurrent oddity found, when modelling stars binaries as single stars \citep[cf.][]{2010A&ARv..18...67T, Higl2017}. It is, of course, a fundamental requirement that the ages of both companions agree within the error bars.


\section{Discussion} \label{sec:discussion}

In this paper, we employ frequency ratios when computing the likelihood. This is common practice \citep[e.g.][]{Aguirre2015}. The reason for this choice is that no viable correction relation exists for the partially corrected frequencies of the coupled models, which renders a match to the individual frequencies pointless for such models. We, therefore, resort to frequency ratios, in order to compare coupled and standard stellar models on an equal footing.

The obtained best-fitting models do robustly recover the desired properties, i.e. the frequency ratios and separations ($\Delta \nu_0$, $\delta \nu_{02}$, $r_{02}$, $r_{10}$, $r_{01}$) as well as spectroscopic constraints ($T_\mathrm{eff}$) within one standard deviation.
However, as a trade-off, we attribute high posterior probabilities to models that do not yield the expected functional dependence of the systematic offset between the model frequencies and observations --- that is, the surface effect (cf. Figs~\ref{fig:Freq_ACA} and \ref{fig:Freq_ACB}). Matching frequency ratios {\color{black}and separations} rather than individual frequencies hence affects the obtained parameter estimates. This is a well-known issue \citep{Aguirre2013}. 
For our purposes, we deem the mentioned trade-off to be acceptable: in this paper, we present a differential study with the aim of comparing two different methods on equal terms. The obtained best-fitting models are meanwhile not assumed to give a perfect physical description of the interior structure of Alpha Centauri A and B. Indeed, {\color{black} we have neglected some physical processes, such as metal diffusion, and have, furthermore, fixed an approximate metallicity.}

As regards the priors, we note that the Gaussian mass prior plays a crucial role, due to the degeneracy between different global stellar parameters. We have thus attempted to restrict the parameters of single stars in the \textit{Kepler} LEGACY sample \citep{Lund2017}, using only uniform mass prior. 
However, we find that the resulting posterior probability distribution is multi-modal, allowing for solutions that have un-physically low or high helium contents. In other words, the likelihood and priors were not sufficiently restrictive.

We repeated the analysis, using older Gaussian mass priors by \cite{Pourbaix2002} and the spectroscopic constraints by \cite{Thevenin2002} for comparison. Doing so, we arrive at the same qualitative conclusions.

We find that the use of coupled and standard stellar models lead to consistent results for the predicted interior structure as well as the global parameters, including the mass, age, and composition. This is consistent with the qualitative conclusions that were drawn by \citetalias{Joergensen2018}, when addressing the present Sun: to this end, as regards the parameter estimates of MS stars, the use of coupled models is not crucial --- nevertheless, a more physically realistic depiction of the surface layers is obtained by using coupling, which is crucial for detailed studies of stellar physics and structures.

However, when using coupled stellar models, the posterior probability of the models becomes nearly insensitive to the mixing length parameter. In other words, coupled models are able to recover the correct stellar properties without suffering from the degeneracy of MLT. By eliminating this degeneracy, the coupling of 1D and 3D models becomes superior to the standard procedure, even for MS stars, yielding more robust parameter estimates.

{We note that $\alpha_\textsc{mlt}$ fulfils a different role in standard stellar models than coupled stellar models:
in standard stellar models, $\alpha_\textsc{mlt}$ must provide the entropy jump between the photosphere and the asymptotic adiabat in the deep stellar interior.
In coupled models, on the other hand, most of this entropy jump takes place in the appended $\langle \mathrm{3D} \rangle$-envelope. MLT is thus only needed to bridge the entropy difference in a nearly-adiabatic layer between the asymptotic adiabat and the matching point. This explains why $\alpha_\textsc{mlt}$ takes a much larger value in coupled stellar models.

Our method for coupling 1D and 3D models is partly motivated by the aim to ensure that the resulting coupled models are continuous at the matching point ({\color{black} cf. Figs \ref{fig:sun_physqua} and \ref{fig:AlphaCenA_struc}}). This is a critical virtue of our method since we employ our method in asteroseismic analyses: discontinuities in the obtained structure leave their fingerprint on seismic properties and greatly complicate such analyses \citepalias[cf.][]{Joergensen2017}. Previously published patching procedures, therefore, likewise, take great care to achieve continuous stratifications \citep[e.g.][]{Trampedach2014b,Ball2016}. 

\begin{figure}
\centering
\includegraphics[width=0.85\linewidth]{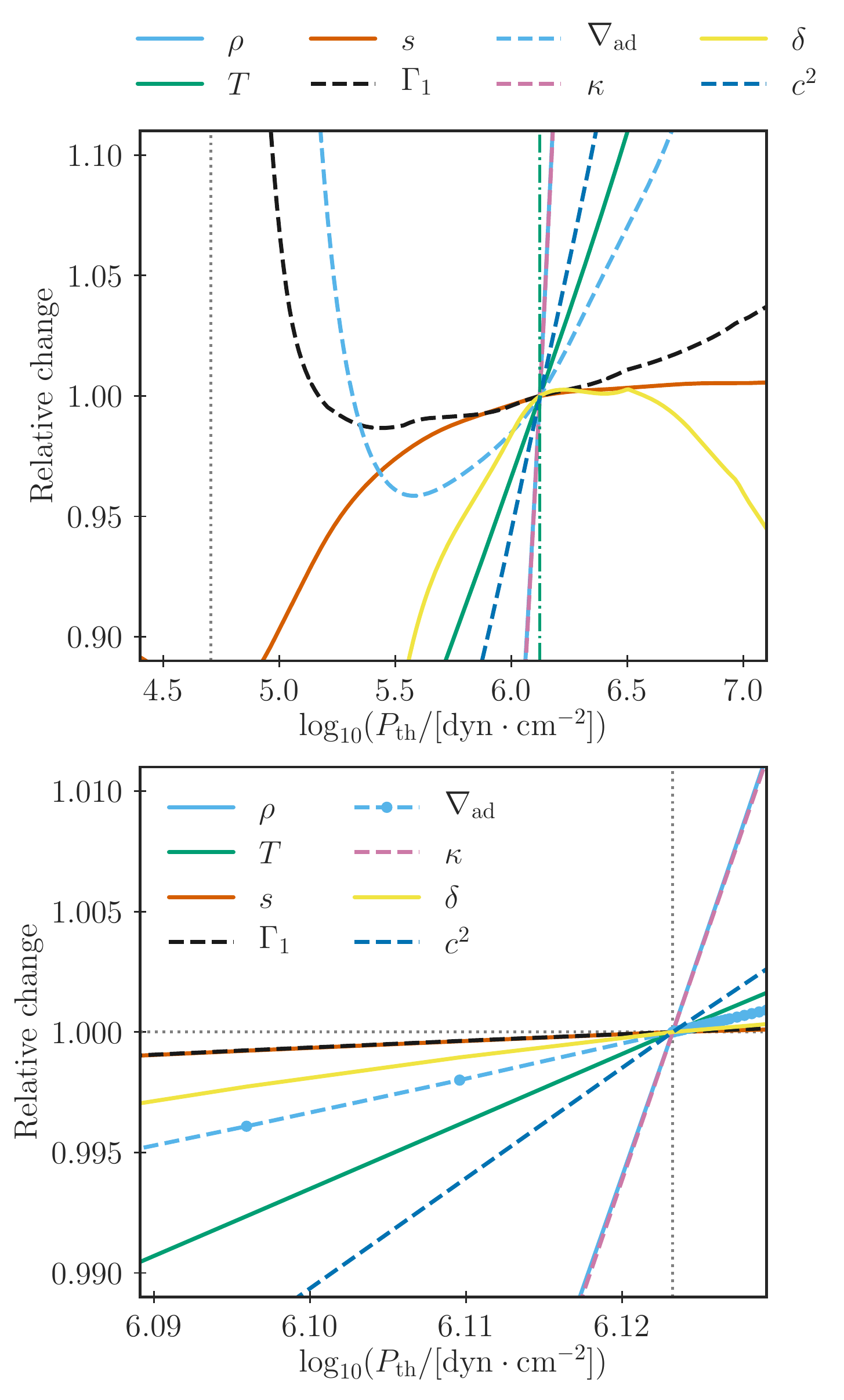}
\caption{\color{black}\textbf{Upper panel:} Structure of the uppermost layers for a coupled model of Alpha Centauri A, showing the change in several physical quantities relative to the corresponding value taken at the matching point as a function of the thermal pressure ($P_\mathrm{th}$). For more details, see the caption of Fig.~\ref{fig:sun_physqua}. \textbf{Lowel panel:} See Fig.~\ref{fig:sun_physqua}.}
\label{fig:AlphaCenA_struc}
\end{figure}

On the other hand, it is worth noting that the global parameters of any 3D simulation uniquely determine its asymptotic adiabat. 
\citet{Ludwig1999} have, therefore, suggested an alternative approach by which $\alpha_\textsc{mlt}$ is calibrated in such a way as for 1D models to recover the correct asymptotic adiabat. However, these models do not recover the stratification of 3D simulations in the nearly-adiabatic and superadiabatic layers. 
Conversely, it stands to reason that neither our method nor that by \citet{Trampedach2014b} is guaranteed to yield the correct deep adiabat. 
While we find that the value of $\alpha_\textsc{mlt}$ has little impact on the best-fitting global parameters of coupled models, one may, therefore, suspect that only certain values of $\alpha_\textsc{mlt}$ make the coupled models physically fully consistent with the underlying 3D simulations.

In order to counter this argument, we have taken a closer look at a subset of standard and coupled models from our Markov chains. We have thus selected a handful of models that mainly differ from the best-fitting model in the associated Markov chain by their value of $\alpha_\textsc{mlt}$: $M$, $X$, $Z$, $\log g$, and $T_\mathrm{eff}$ match the corresponding values of the best-fitting model within $10^{-2}M_\star$, $2\times10^{-4}$, $10^{-4}$, $10^{-3}\,$dex, and $5\,$K, respectively --- here, $M_\star$ denotes the mass of the corresponding best-fitting model. For each of these models, we have evaluated the entropy at a thermal pressure of $10^{11}\,\mathrm{dyn}\,\mathrm{cm^{-2}}$, i.e. deep within the adiabatically convective region. At this depth, we have reached the adiabat of the model.
For simplicity, we compute the entropy from the temperature and pressure by assuming that the gas behaves like an ideal gas, i.e. that the level of ionization is constant --- this is a very reasonable approximation at the considered depth:
\begin{equation}
s=\frac{k_\mathrm{B}}{\mu m_\mathrm{u}} \ln \left( \frac{T^{5/2}}{P_\mathrm{th}} \right), \label{eq:entropy_ideal}
\end{equation}
where $m_\mathrm{u}$ is the atomic mass unit, $k_\mathrm{B}$ is the Boltzmann constant, and $\mu$ is the mean molecular weight.

As expected, we find the evaluated adiabat of the standard models to be rather sensitive to the exact value of the mixing length parameter: by varying $\alpha_\textsc{mlt}$ from 1.75 to 1.80, we shift the entropy of the deep adiabat by roughly 2 per cent. In order to achieve a comparable shift of the deep adiabat in the coupled models, $\alpha_\textsc{mlt}$ must be doubled. This result is illustrated in Fig.~\ref{fig:SA_1D3D}. It emphasizes the conclusion drawn from the posterior probability distributions: coupled models are virtually insensitive to the value of the mixing length parameter. In other words, since $\alpha_\textsc{mlt}$ only affect the stratification of a thin nearly-adiabatic layer, large changes in $\alpha_\textsc{mlt}$ are required for these changes to have any significant effect on the asymptotic adiabat.

\begin{figure}
\centering
\includegraphics[width=0.85\linewidth]{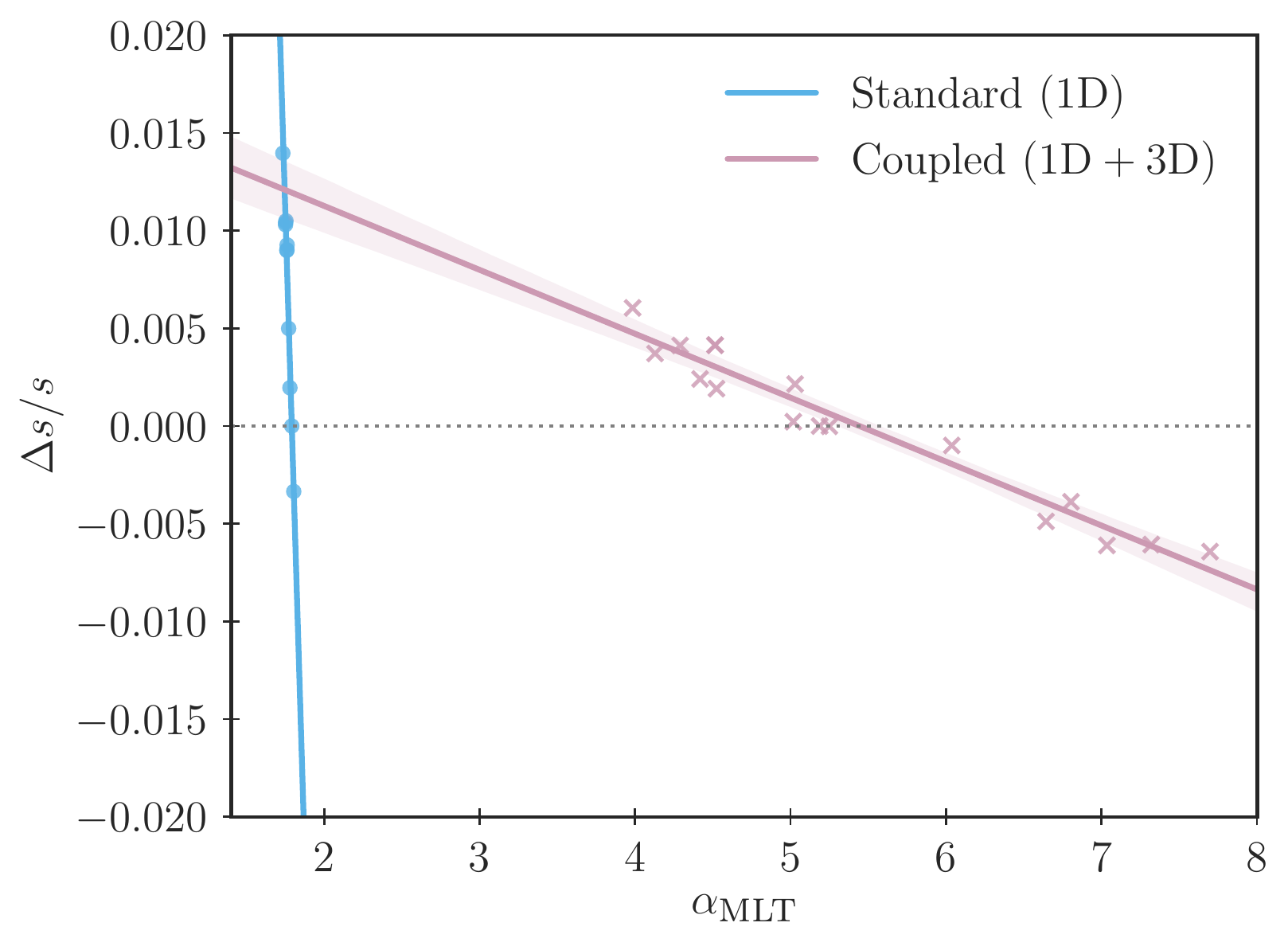}
\caption{Entropy difference between a subset of models from each of the Markov chain for Alpha Centauri A and the corresponding best-fitting models as a function of the mixing length parameter ($\alpha_\textsc{mlt}$). The only notable difference between the models in the each subset is the value of the mixing length parameter; all other global parameters are identical within an accuracy specified in the text.
}
\label{fig:SA_1D3D}
\end{figure}

The fact that the interior structure is rather insensitive to $\alpha_\textsc{mlt}$ is an intriguing feature of our coupled models since it implies that the evolution tracks are likewise insensitive to this parameter. As briefly mentioned \citetalias{Joergensen2018}, this is indeed the case. We can, therefore, draw sound conclusions about the RGB, despite the fact that we are keeping $\alpha_\textsc{mlt}$ constant throughout the evolution. This is not the case for standard stellar models: as shown by \cite{Trampedach2014b, Tayar2017}, the mixing length parameter of such models has to vary across the HR diagram, in order to recover observations or the results from multi-dimensional hydrodynamic simulations. In our coupled models, on the other hand, the interior structure is dictated by the appended $\langle \mathrm{3D} \rangle$-envelope rather than by $\alpha_\textsc{mlt}$.
This point is also illustrated in Fig.~\ref{fig:chal_1D3D}, where we have plotted stellar evolution tracks assuming different $\alpha_\textsc{mlt}$: all models take the corresponding best-fitting models as their starting point but use different $\alpha_\textsc{mlt}$ in the subsequent evolution. For the coupled models, $\alpha_\textsc{mlt}$ varies between 4.0 and 9.0, i.e. by more than a factor of 2. For the standard stellar models, $\alpha_\textsc{mlt}$ varies from 1.70 to 1.95, i.e. by roughly 15 per cent. As can be seen from the figure, the evolution tracks of the coupled models lie closer together than those of the standard stellar models, despite the substantially larger variation in $\alpha_\textsc{mlt}$.

\begin{figure}
\centering
\includegraphics[width=0.85\linewidth]{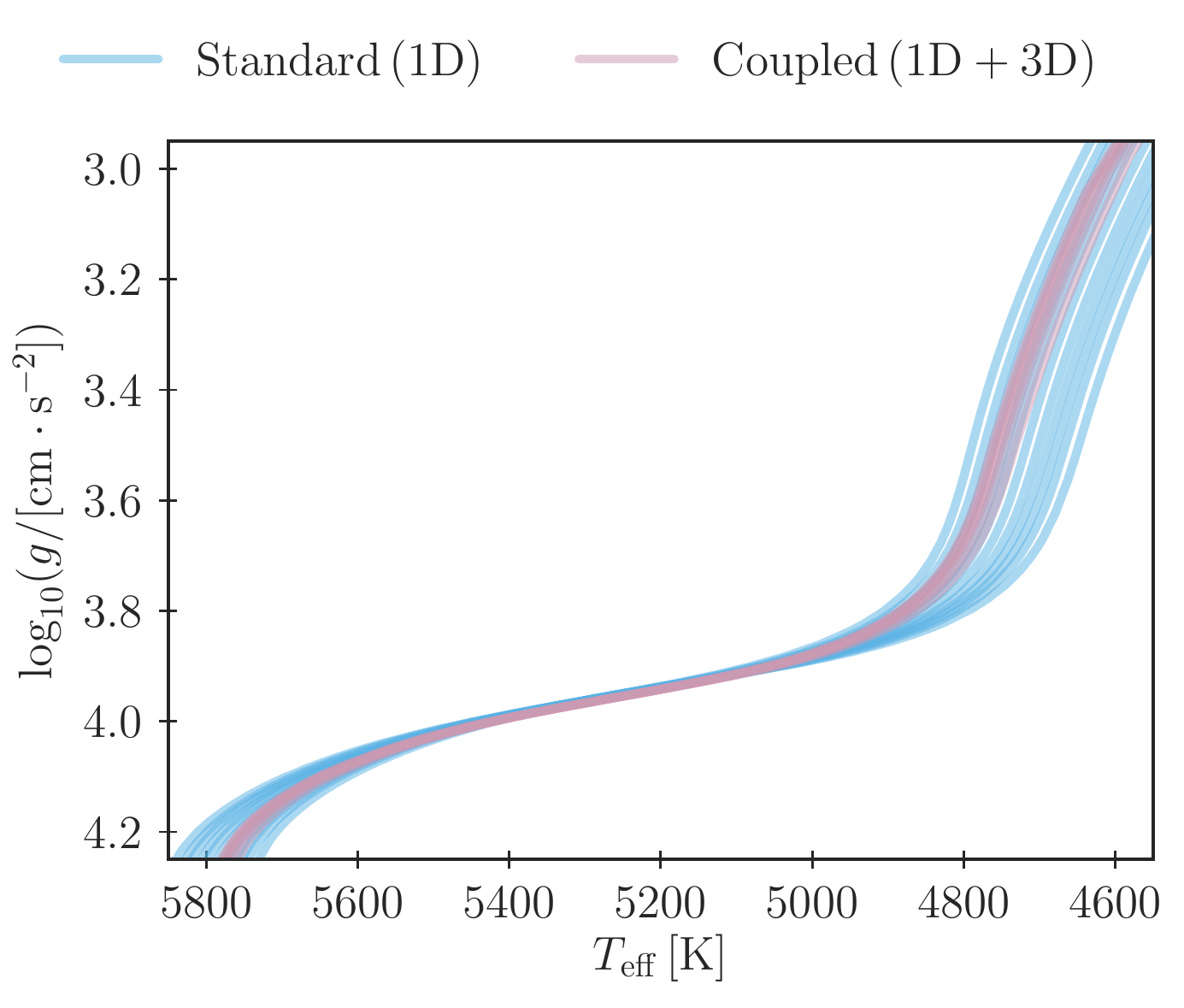}
\caption{Stellar evolution tracks for the evolution of Alpha Centauri A beyond its current evolutionary state, assuming different values for $\alpha_\textsc{mlt}$. For the standard stellar models, $\alpha_\textsc{mlt}$ varies between 1.70 and 1.95. For the coupled stellar models, $\alpha_\textsc{mlt}$ varies between 4.0 and 9.0. Note that the evolution tracks of the coupled models cover several of the evolution tracks of the standard stellar models. {\color{black}The standard stellar models show a larger deviation in the evolution tracks on the MS as well as on the RGB, due to the high sensitivity to $\alpha_\textsc{mlt}$.} On the RGB, at $\log g = 3.5$, the effective temperatures of the standard stellar models deviate by roughly $140\,$K, while the coupled stellar models only deviate by roughly $30\,$K. 
}
\label{fig:chal_1D3D}
\end{figure}

{\color{black}As regards Fig.~\ref{fig:chal_1D3D}, we note that the standard stellar models do not only show a larger spread on the RGB but also on the MS. 
Both for the RGB and the MS, the spread is due to the fact that we solely vary $\alpha_\textsc{mlt}$ when performing the comparison. In order to avoid the spread on the MS, one would need to adjust other physical properties of the stellar models, including its mass or its chemical composition. In other words, in the case of standard stellar models, $\alpha_\textsc{mlt}$ is highly degenerate with other global parameters. The evolution tracks of coupled stellar models, on the other hand, are practically insensitive to $\alpha_\textsc{mlt}$, on the MS as well as on the RGB, which implies that coupled models do not suffer from the same degeneracy.}

{\color{black}Moreover, as regards our choice only to vary $\alpha_\textsc{mlt}$, we note that any changes in the global parameters, such as the mass or the chemical composition, would mean that we would no longer be comparing the same star and internal structures, rendering any such comparison meaningless.
We furthermore note that the evolution on the RGB exhibits no memory of $\alpha_\textsc{mlt}$ and hence of $T_\mathrm{eff}$ on the MS. In other words, we would achieve the same spread on the RGB, whether we were to change
$\alpha_\textsc{mlt}$ from the zero age main-sequence or during the subgiant branch and onwards.}

As mentioned above, the asymptotic adiabat of any given 3D simulation is uniquely determined by its global parameters. Until now, we have argued that the value of $\alpha_\textsc{mlt}$ has little influence on the achieved adiabat. Whether the correct adiabat is obtained, is a different question. As shown in Fig.~2 of 
\citetalias{Joergensen2018} based on a solar calibration, however, coupled models turn out to closely recover the temperature of the underlying 3D simulation as a function of pressure \textit{below} the matching point --- this comparison can be made in the case of the Sun, since a Stagger-grid simulation at this point of the HR diagram exists. In other words, the deep interiors of coupled models closely match the asymptotic adiabat that would make them consistent with the underlying 3D simulations.

Since our coupled models ensure a continuous stratification at the matching point, deeper 3D simulations would make the models converge towards a unique adiabat and render $\alpha_\textsc{mlt}$ completely obsolete. Grids containing such 3D simulations are, however, currently not available.}

In summary, the coupling method by \citetalias{Joergensen2018} provides a viable alternative to MLT: not only does coupling yield a more realistic depiction of superadiabatic convection, but it also reduces the importance and influence of $\alpha_\mathrm{mlt}$, making the models almost insensitive to it.
This is a compelling feature of coupled models since $\alpha_\mathrm{mlt}$ is degenerate with other stellar parameters, including the helium abundance and the mass of the star. Furthermore, it is still not settled how to vary the mixing length parameter across the HR diagram correctly and throughout the envelope \citep[e.g.][]{Schlattl1997,Tayar2017}.


\section{Conclusions}

{\color{black}Using Bayesian inference, we present a differential study of Alpha Centauri A and B, in order to demonstrate how stellar models may benefit from a physically more realistic depiction of the outermost layers of stars with convective envelopes. For this purpose, we compare the stellar parameter estimates obtained using two different descriptions of superadiabatic convection: while one set of models draws on MLT to describe the surface layers, the other set of models reliably mimics the stratification of 3D radiative hydrodynamic simulations of convection based on the coupling scheme by \citetalias{Joergensen2018}.}

By choosing Alpha Centauri A and B, we provide the first analysis based on coupled models for stars with non-solar metallicity and hereby further demonstrate the efficacy of the coupling scheme by \citetalias{Joergensen2018}. Moreover, owing to their proximity and binary nature, Alpha Centauri A and B are well-studied benchmarks for stellar evolution.

We demonstrate that the deep adiabat and the evolution tracks of coupled models are rather insensitive to the value of the mixing length parameter. Likewise, the mixing length parameter plays no major role for the established parameter estimates.
Thus, the coupling of 1D and 3D models provides a viable improvement of stellar models and a superior alternative to MLT: not only does coupling yield a more realistic depiction of superadiabatic convection, but it also reduces the importance and influence of $\alpha_\mathrm{mlt}$, making the models almost insensitive to this parameter.
This is a compelling feature of coupled models since $\alpha_\mathrm{mlt}$ is degenerate with other stellar parameters, including the helium abundance and the mass of the star.

As discussed in this paper, the change to a more realistic description of the boundary layers affects the predicted stellar evolution. 
One needs to look no further than \citet{1955ApJ...121..776H} to understand the importance of the correct boundary conditions for the stellar structure equations.
While the inferred global parameters for main-sequence stars are impacted minimally, it is not clear that this is the case for later evolutionary phases {\color{black}with deeper convective envelopes.}
The obtained best-fitting models thus follow a different evolutionary paths. We aim to quantify this for later stages in future work but note that there are several issues that we must first overcome: the high computational cost of using a MCMC, the increased computational effort of computing stars at later evolutionary stages, and the low resolution of the Stagger grid at low $\log g$.
Furthermore, by choosing Alpha Centauri A and B, we present a study of stars {\color{black}whose metallicities are not too far from solar}. Our conclusion should hence be validated, by addressing MS stars with vastly different compositions. This is beyond the scope of the present paper.

{\color{black}While we are merely performing a differential study, and while the obtained best-fitting models may by no means yield a complete description of Alpha Centauri A and B, our analysis of this binary clearly illustrates the vital role that more realistic outer boundary layers play for our understanding of stellar evolution.
Continuing this endeavour by characterizing a broad range of stars thus promises to be a fruitful endeavour.}


\section*{Acknowledgements}

First and foremost, we would like to record our sincere gratitude to Achim Weiss for his guidance and input. His valuable insights and our fruitful discussions have greatly contributed to this paper.
We would like to thank Zazralt Magic for providing the Stagger grid as well as James Kuszlewicz and Earl Bellinger for their useful discussions. 

Part of the research leading to the presented results has received funding from the European Research Council under the European Community's Seventh Framework Programme (FP7/2007-2013) / ERC grant agreement no 338251 (StellarAges).

\emph{Software:} Calculations were performed using the \textsc{hephaestus} pipeline (Angelou, in prep) making use of the \textsc{garstec} \citep{Weiss2008,Joergensen2018} stellar evolution code and the \textsc{ADIPLS} pulsation package 0.3 \citep{Christensen-Dalsgaard2008a}. 
Analysis in this manuscript was performed with  anaconda 3.6.8 and python 3.6.8 packages NumPy 1.15.0 \citep{van2011numpy}, scipy 1.2.1 \citep{scipy}, pandas 0.23.3 \citep{mckinney2010data}, emcee 2.2.1 \citep{emcee}, matplotlib \citep{matplotlib}, corner 2.0.1 \citep{corner}, wquantiles 0.5 \citep{quantiles} and sobol \citep{sobol}.



\bibliographystyle{mnras}
\bibliography{manual_refs,mendeley_export}








\bsp    
\label{lastpage}
\end{document}